\documentclass[pra,twocolumn,superscriptaddress,loatfix,nofootinbib,amsmath,amssymb,longbibliography]{revtex4-1}

\usepackage[english]{babel}
\usepackage[T1]{fontenc}
\usepackage{float}
\usepackage{graphicx}
\usepackage{bm}
\usepackage{braket}
\usepackage{xcolor}
\usepackage{amsfonts}
\usepackage{comment}
\usepackage{dirtytalk}
\usepackage{hyperref}
\usepackage{cleveref}
\usepackage{url}
\usepackage{bbold}
\usepackage{realboxes}

\usepackage{enumitem}
\usepackage{algpseudocode}
\usepackage{listings}
\usepackage{textcomp}

\usepackage[caption=false]{subfig}

\definecolor{bkgd}{RGB}{240,242,246}
\definecolor{ceruleanblue}{rgb}{0.16, 0.32, 0.75}
\definecolor{orange-red}{rgb}{1.0, 0.27, 0.0}
\definecolor{anotherblue}{RGB}{37,92,243}
\definecolor{blackblue}{RGB}{46,60,85}
\definecolor{goldyellow}{RGB}{199,146,12}
\lstdefinestyle{altstyle2}{
    backgroundcolor=\color{bkgd},
    basicstyle=\ttfamily\footnotesize\color{blackblue},
    breakatwhitespace=false,
    breaklines=true,
    captionpos=b,
    commentstyle=\color{goldyellow},
    keepspaces=true,
    keywordstyle=\color{orange-red},
    language=Python,
    numbersep=5pt,
    numberstyle=\tiny\color{ceruleanblue},
    showspaces=false,
    showstringspaces=false,
    showtabs=false,
    stringstyle=\color{anotherblue},
    tabsize=2
}
\usepackage[utf8]{inputenc}
\usepackage{algorithm}
\usepackage{xparse}
\usepackage[margin=1in]{geometry}
\usepackage{titlesec}

\def \l {\left}
\def \r {\right}
\newcommand{\twoMat}[4]{\l( \begin{array}{cc} #1 & #2 \\ #3 & #4 \end{array} \r) }

\begin{document}

\title{\Large{Simulations of Quantum Circuits with Approximate Noise \\ using qsim and Cirq}}
\author{Sergei V. \surname{Isakov}}
\affiliation{Google}

\author{Dvir Kafri}
\affiliation{Google}

\author{Orion Martin}
\affiliation{Google}

\author{Catherine \surname{Vollgraff Heidweiller}}
\affiliation{Google}

\author{Wojciech Mruczkiewicz}
\affiliation{Google}

\author{Matthew P. \surname{Harrigan}}
\affiliation{Google}

\author{Nicholas C. Rubin}
\affiliation{Google}

\author{Ross Thomson}
\affiliation{Google}

\author{Michael Broughton}
\affiliation{Google}

\author{Kevin Kissell}
\affiliation{Google}

\author{Evan Peters}
\affiliation{Fermi National Accelerator Laboratory}

\author{Erik Gustafson}
\affiliation{Fermi National Accelerator Laboratory}

\author{Andy C. Y. Li}
\affiliation{Fermi National Accelerator Laboratory}

\author{Henry Lamm}
\affiliation{Fermi National Accelerator Laboratory}

\author{Gabriel Perdue}
\affiliation{Fermi National Accelerator Laboratory}

\author{Alan K. Ho}
\affiliation{Google}

\author{Doug Strain}
\affiliation{Google}

\author{Sergio Boixo }
\affiliation{Google}

\date{November 4th, 2021}

\begin{abstract} 
We introduce multinode quantum trajectory simulations with qsim, an open source high performance simulator of quantum circuits. qsim can be used as a backend of Cirq, a Python software library for writing quantum circuits. We present a novel delayed inner product algorithm for quantum trajectories which can result in an order of magnitude speedup for low noise simulation. We also provide tools to use this framework in Google Cloud Platform, with high performance virtual machines in a single mode or multinode setting. Multinode configurations are well suited to simulate noisy quantum circuits with quantum trajectories. Finally, we introduce an approximate noise model for Google's experimental quantum computing platform and compare the results of noisy simulations with experiments for several quantum algorithms on Google's Quantum Computing Service. 
\end{abstract}

\maketitle

\tableofcontents

\section{Introduction}

Classical software which simulates quantum circuits with an approximate noise model of quantum hardware enables the study of NISQ quantum algorithms and applications discovery. qsim~\cite{quantum_ai_team_and_collaborators_2020_4023103,isakov_blog_2020} was recently launched to allow users of Google Quantum AI open source ecosystem of software tools to simulate quantum circuits more efficiently on classical processors. These software tools include Cirq~\cite{cirq_developers_2021_5182845}, a quantum programming framework, ReCirq~\cite{quantum_ai_team_and_collaborators_2020_4091470}, a repository of research examples, and application-specific libraries such as OpenFermion~\cite{mcclean2020openfermion} for quantum chemistry and TensorFlow Quantum~\cite{broughton2020tensorflow} for quantum machine learning. New features have been added to qsim and Cirq to make quantum circuit simulations more performant and intuitive, and to make noise simulations more sophisticated.

In this  paper, we describe the theory and software routines which underpin qsim's performance.
We outline qsim implementations and workflows for various classical processor types and setups, including single and multinode CPU and GPU setups. Finally, we describe a generic noise model which approximates Google's Quantum Computing Service (QCS).

\section{Cirq: a programming framework for quantum circuits}
Cirq~\cite{cirq_developers_2021_5182845} is a Python software library for writing, manipulating, optimizing and running quantum circuits on quantum computers and quantum simulators. 
Cirq can be used with experimental quantum processors, such as Google's Quantum Computing Service, Alpine, Pasqal, Rigetti and IonQ. It comes with built-in Python simulators for testing small circuits, and supports high performance simulators, such as Qulacs~\cite{suzuki2021qulacs} and quimb~\cite{gray2018quimb}. Cirq has also been integrated with other software libraries, such as QC Ware Forge, Xanadu Pennylane~\cite{bergholm2018pennylane}, Zapata Orquestra, Sandia National Lab pyGSTi~\cite{nielsen2020probing}, CQC t|ket>~\cite{sivarajah2020t} and Quantum Benchmark True-Q~\cite{beale_stefanie_j_2020_3945250}. Cirq is part of Google Quantum AI open source ecosystem, which includes ReCirq~\cite{quantum_ai_team_and_collaborators_2020_4091470}, OpenFermion~\cite{mcclean2020openfermion} and TensorFlow Quantum~\cite{broughton2020tensorflow}.  In this paper, we will focus on the use of Cirq to simulate approximate experimental noise with qsim~\cite{quantum_ai_team_and_collaborators_2020_4023103,isakov_blog_2020}.

A \lstinline{Qubit} in Cirq is an abstract object that has an identifier. The actual state of a qubit or qubits is maintained in a quantum processor or a simulator. A \lstinline{Gate} in Cirq is an effect that can be applied to a collection of qubits. A \lstinline{Gate} can be a unitary gate or a quantum channel. Quantum channels can represent noise, such as amplitude or phase damping channels.  

The primary representation of quantum programs in Cirq is the \lstinline{Circuit} class. A \lstinline{Circuit} is a collection of \lstinline{Moment}s. Each \lstinline{Moment} is a collection of \lstinline{Operation}s that all act during the same time slice, but in different qubits. An \lstinline{Operation} in Cirq is a \lstinline{Gate} that has been applied to qubits.
\begin{lstlisting}
import cirq

q0, q1, q2 = cirq.LineQubit.range(3)

moment0 = cirq.Moment([
    cirq.CZ(q0, q1), cirq.X(q2)
])
moment1 = cirq.Moment([cirq.CZ(q1, q2)])
circuit = cirq.Circuit((moment0, moment1))
\end{lstlisting}

Cirq also includes tools to transform circuits, including adding quantum channels after unitary gates in  a circuit to simulate noisy experimental quantum processors. Noise can be added to a Cirq circuit before constructing a simulator object and simulating the circuit. There are two procedures for adding noise to a Cirq circuit: adding individual noise events, or defining a global noise model.

An individual noise event can be added as a \lstinline{cirq.Channel} with corresponding noise parameters to the Cirq circuit (in the \lstinline{cirq.Circuit} argument). Calling the \lstinline{cirq.kraus} protocol on a channel returns the Kraus operators corresponding to that channel.
All channels are subclasses of \lstinline{cirq.Gate}. As such, they can act on qubits and be used in circuits in the same manner as gates. 

Cirq has multiple common noise channel options built in, such as the depolarizing channel \lstinline{cirq.depolarize}, the phase damping channel \lstinline{cirq.phase_damp} and the bit flip channel \lstinline{cirq.bit_flip}. Channels can be controlled by appending \lstinline{.controlled}. Custom channels can be defined using \lstinline{MixedUnitaryChannel} or \lstinline{KrausChannel}. \lstinline{MixedUnitaryChannel} takes a list of (probability, unitary) tuples and uses it to define the \lstinline{_mixture_} method. \lstinline{KrausChannel} takes a list of Kraus operators and uses it to define the \lstinline{_kraus_} method. A measurement key can be used as a parameter in a custom noise channel. This key will be used to store the index of the selected unitary or Kraus operator in the measurement results.

Noise that affects an entire circuit can be described with the \lstinline{cirq.NoiseModel} type. Objects of this type must define one of three methods to describe how to convert a ``clean'' circuit into a ``noisy'' circuit: (1) \lstinline{noisy_operation}, which mutates each operation independently, (2) \lstinline{noisy_moment}, which mutates each set of simultaneous operations, or ``moment'', as a group, or (3) \lstinline{noisy_moments}, which mutates the entire circuit at once.
A simple version of this is provided in \lstinline{cirq.ConstantQubitNoiseModel}, which applies a specified gate or channel to every qubit in the circuit at the start of each moment. For more complex behavior, users can implement their own \lstinline{NoiseModel} type. 

Once constructed, a \lstinline{NoiseModel} can be applied to a circuit using the \lstinline{Circuit.with_noise} method. This generates a ``noisy'' version of the original circuit, which can then be simulated with qsim or one of the builtin Cirq simulators.

\section{qsim: a quantum circuit simulator}
qsim is a full state vector quantum circuit simulator: it computes all the $2^n$ amplitudes of the state-vector, where $n$ is the number of qubits.  Essentially, in order to apply gates or operators to the state vector, the simulator performs matrix-vector multiplications repeatedly.  We use single precision arithmetic and gate fusion~\cite{smelyanskiy2016,haener2017} to speed up the simulation.  In addition, we use SIMD (single instruction/multiple data) instructions for vectorization and OpenMP for multi-threading on CPUs.  Three SIMD versions are available: SSE, AVX2/FMA, and AVX512.  We also have a GPU implementation using CUDA.

\subsection{Matrix-vector multiplication}

For a $q$-qubit gate and an $n$-qubit state vector, the full matrix-vector multiplication can be block diagonalized into $2^{n-q}$ (gate matrix)-subvector multiplications, where the gate matrix is of size $2^q \times 2^q$ and each subvector is of size $2^q$ as depicted in Algorithm~\ref{alg:mvmult}.

\begin{algorithm}[H] 
\caption{Matrix-vector multiplication algorithm.}
\label{alg:mvmult}
\begin{algorithmic}[1]
\For {$i \gets 0,2^{n-q}-1$}
  \State $M \gets 2^{q}$
  \State $w$ \Comment{temporary vector of size $M$}
  \State $v \gets $ subvector(state, gate\_qubits) \Comment{read subvector from memory}
  \For {$j \gets 0,M-1$}
    \State $w_j \gets \sum_{k=0}^{M-1} U_{jk} v_k $
  \EndFor
  \State subvector(state, gate\_qubits) $\gets w$  \Comment{write results back to memory}
\EndFor
\end{algorithmic}
\end{algorithm}

The total number of flops is approximately equal to $2^{q+2}\cdot 2^{n+1}$ and the total number of bytes to read and write is $2^{n+4}$ (single precision).  The arithmetic intensity (the ratio of the number of flops to the number of bytes to read and write) is $2^{q-1}$.  The arithmetic intensity is small for small values of $q$ and the performance is usually limited by the memory bandwidth.  It is beneficial to fuse small gates into larger gates to increase the arithmetic intensity and better utilize the compute power of modern CPUs and GPUs~\cite{smelyanskiy2016,haener2017}.  

\subsection{Gate fusion}\label{sec:fusion}

A quantum circuit can be considered as a lattice structure that has spatial and time directions. The time direction corresponds to the order in which gates are applied.  The core idea of gate fusion is to combine gates that are close in space and time into larger gates~\cite{smelyanskiy2016,haener2017}.  Gate fusion increases the arithmetic intensity, decreases the number of gates, and typically leads to a significant speedup.

There are two steps in the fusion algorithm that is employed in qsim.  First, large gates and small gates that are neighbors in time and act on the same qubits are combined (say, 2-qubit gates are combined with 1-qubit gates).  Second, the algorithm greedily combines gates that are close in space and time.  Essentially, this works as follows.  All the resulting gates from the first step are unmarked.  The unmarked gates are picked for processing in increasing time order.  The first unmarked gate is picked and marked.  The nearest (unmarked) neighbors (the gates that share qubits with the picked gate) forth in time and the next nearest unmarked neighbors back in time (if they do not have unmarked neighbors further back in time) are added while the resulting fused gate is no greater than the specified maximum fuse size $f$.  All added gates are marked.  This procedure is repeated until all the unmarked gates are exhausted.

Typically the optimal value of the maximum fuse size $f$ is 4 for large numbers of threads (or on GPUs) and large circuits.  Smaller fuse size, $f=2$ or $f=3$, can be optimal for small numbers of threads and/or small circuits.  However, this might depend on the circuit structure and the user is advised to try out different values.

\subsection{SIMD implementation}

We use SIMD instructions (single instruction/multiple data) to make the most of the compute power of CPUs.  We avoid the usage of horizontal SIMD instructions by keeping the real and imaginary parts of $k$ state-vector amplitudes in separate SIMD registers, where $k$ is the SIMD register size in floats.  In single precision, $k=4$ for SSE, $k=8$ for AVX, and $k=16$ for AVX512.  The real and imaginary parts can be stored in memory separately in blocks of size $k$ ($k$ real parts are followed by $k$ imaginary parts and so on) or alternatively they can be stored conventionally (one real part is followed by one imaginary part and so on).  In the former case, the real and imaginary parts can be loaded immediately into two SIMD registers.  In the latter case, the amplitudes have to be reshuffled.  This technique allows us to perform the (gate matrix)-subvector multiplications in parallel for up to $k$ subvectors.

We denote the qubit indices that are larger than or equal to $\log_2(k)$ as ``high'' qubit indices and the qubit indices that are smaller than $\log_2(k)$ as ``low'' qubit indices.  It is straightforward to generalize Algorithm~\ref{alg:mvmult} to make use of SIMD instructions if all the gate qubits are high.  In this case, the first loop runs from 0 to $2^{n-q-\log_2(k)}-1$.  $k 2^q$ state-vector amplitudes are loaded into $2\cdot 2^q$ SIMD registers (the additional factor of two is because of real and imaginary parts) and SIMD arithmetic instructions are used to calculate $k$ matrix-vector products simultaneously.

If some of the gate qubits (or all) are low then the algorithm is more involved.  Let $h$ be the number of high qubits and $l$ a number of low qubits ($h+l=q$).  Now the first loop in Algorithm~\ref{alg:mvmult} runs from 0 to $2^{n-h-\log_2(k)}-1$ and the second loop runs from 0 to $2^h-1$.  $k 2^h$ state-vector amplitudes are loaded into $2\cdot 2^h$ SIMD registers.  We calculate $k/2^l$ matrix-vector products simultaneously.  To do that, $k 2^h$ state-vector amplitudes are rearranged into $2\cdot(2^q-2^h)$ additional SIMD registers ($2^l-1$ additional registers for each register loaded from memory).  We use the following specific permutations
$$
  p_{k,i} = (i \oplus k) | (i \& \tilde{m}).
$$
Here the $i$th data element of the register loaded from memory goes to the position $p_{k,i}$ into the $k$th additional register ($k$ runs from 1 to $2^l-1$), $m$ is the low qubit binary mask, $\tilde{m}$ is the one's complement of $m$ and $\oplus$ denotes the summation of masked bits.  An example for AVX ($k=8$) and $l=2$ with low qubit indices equal to 0 and 1 ($m=3$) follows. Each register loaded from memory
\begin{eqnarray*}
  &&a_7a_6a_5a_4a_3a_2a_1a_0
\end{eqnarray*}
is rearranged into three additional registers as
\begin{eqnarray*}
  &&a_4a_7a_6a_5a_0a_3a_2a_1\\
  &&a_5a_4a_7a_6a_1a_0a_3a_2\\
  &&a_6a_5a_4a_7a_2a_1a_0a_3.
\end{eqnarray*}
The matrix elements to calculate the sum in line 6 in Algorithm~\ref{alg:mvmult} are loaded into SIMD registers accordingly.  $\bar{U}_{jk,i} = U_{j'k'}$, where $\bar{U}_{jk,i}$ denotes the matrix element that is loaded into $i$th position in the $(jk)$th SIMD register, $j$ runs from 0 to $2^h-1$, $k$ runs from 0 to $2^q-1$, $j'=s / 2^q$, $k'=s \bmod 2^q$ and s is given by $j 2^{l+q} + r_i 2^q + 2^l (k / 2^l) + (r_i + k) \bmod 2^l$, where $r_i$ is equal to $i$ compressed with respect to $m$.

\subsection{GPU implementation}

The GPU implementation is very similar to the SIMD implementation.  The GPU implementation uses CUDA.  32-thread warps are used instead of SIMD registers and SIMD instructions, i.e. $k=32$ in this case.  A single thread in a warp basically performs the same role as a single data element in a SIMD register.  $k/2^l$ matrix-vector products are calculated in parallel by a single warp, where $l$ is the number of low qubits.  The implementation efficiently utilizes the GPU compute resources and memory bandwidth, especially, in the case of $l=0$.

We compare the runtime between the CPU and GPU implementations for several CPU and GPU families available in Google Cloud Platform in Sec.~\ref{sec:qsim_runtime} (see Figs.~\ref{fig:qubits_vs_time} and~\ref{fig:perf_noise}).

\subsection{Quantum trajectories in qsim} \label{sec:trajectory_sim_section}

qsim supports noisy circuit simulations using quantum trajectories. A quantum trajectory is implemented by choosing one Kraus operator $K_i$ for each quantum channel with Kraus operators $\{K_i\}$. The probability to sample the Kraus operator $K_i$ is $p_i =\langle \Psi| K^\dagger_i K_i | \Psi \rangle$.  The probabilities $p_i$ sum to unity for each channel, $\sum p_i = 1$. A Kraus operator is sampled and applied to the state vector for each channel sequentially.  This procedure is typically repeated many times, once per quantum trajectory. The Monte Carlo statistical error for an observable estimated with quantum trajectories goes like $1/\sqrt {r}$, where $r$ is the number of trajectories. Therefore the number of trajectories is typically in the thousands or higher. 

We present a novel delayed inner product algorithm for quantum trajectories which can result in an order of magnitude speedup for low noise simulation. In the conventional quantum trajectory algorithm, at least one Kraus operator is applied immediately for each channel. The corresponding probability $p_i$, the norm of the resulting state vector, is calculated. Sometimes multiple probabilities have to be calculated for the same quantum channel to sample with probabilities $\{p_i\}$. 

Our improved delayed inner product algorithm uses a lower bound $\bar{p}_i$ for each sampling probability $p_i$. The lower bound $\bar{p}_i$ is given by the smallest singular value (squared) of the operator matrix $K_i$.  Note that in general, the bounds $\bar{p}_i$ sum to a value $s$ that is smaller than unity.  To sample the Kraus operator, we draw a random number $r$ from the range $[0,1)$.  If $r < s$ then there is no need to apply any Kraus operator immediately and we avoid computing any inner product $\langle \Psi| K^\dagger_i K_i | \Psi \rangle$.  The operator can be sampled just by using the lower bounds. In this case, the application of the picked operator can be deferred.  Deferring the operator application allows us to make use of gate fusion, see Sec.~\ref{sec:fusion}.  If $r \geq s$ then, first, we fuse and apply all the operators that were deferred in the previous steps.  Second, we use the conventional sampling procedure. The algorithm is depicted in Algorithm~\ref{alg:qtrajectory}.  Note that the expectation value in step 14 can be calculated in place without copying the state vector to a temporary vector.  This reduces the memory usage.

Note that, in the case of weak noise, the sum $s$ of lower bounds is typically close to one and the operators get deferred with a high probability.  This gives rise to a significant speedup.  We observe that the runtime is linear in noise strength for typical noise values, see Fig.~\ref{fig:perf_noise}. The runtime of the conventional algorithm is weakly dependent on the noise probability, so the runtime at a noise probability of 0.1 gives a lower bound for the runtime of the conventional algorithm. We observe an order of magnitude runtime speedup for low noise with 27 qubits. Furthermore, if all Kraus operators in a channel are proportional to unitary matrices (like in the depolarizing channel or other mixtures of unitaries), then $s=1$ and we always defer the application of such a channel.

\begin{algorithm}[H]
\caption{Quantum trajectory algorithm.}
\label{alg:qtrajectory}
\begin{algorithmic}[1]
\ForAll{channels}
  \State $k \gets $ size(channel)
  \State $r \gets $ random([0,1))
  \For {$i \gets 1,k$} \Comment{iterate over Kraus operators}
    \If{$r < \bar{p}_i$}
      \State defer applying $K_i$ \Comment{pick $i$th Kraus operator}
      \State go to next channel
    \Else
      \State $r \gets r - \bar{p}_i$
    \EndIf
  \EndFor
  \State fuse and apply all deferred Kraus operators
  \For {$i \gets 1,k$} \Comment{iterate over Kraus operators}
    \State $p_i \gets \langle \Psi| K^\dagger_i K_i | \Psi \rangle$
    \If{$r < (p_i - \bar{p}_i)$}
      \State $|\Psi \rangle \gets (1/\sqrt{p_i}) K_i |\Psi \rangle$ \Comment{pick $i$th Kraus operator}
      \State go to next channel
    \Else
      \State $r \gets r - (p_i - \bar{p}_i)$
    \EndIf
  \EndFor
\EndFor
\end{algorithmic}
\end{algorithm}

\subsection{Main qsim runtime factors}\label{sec:qsim_runtime}

\begin{figure}
    \centering
    \includegraphics[width=\linewidth]{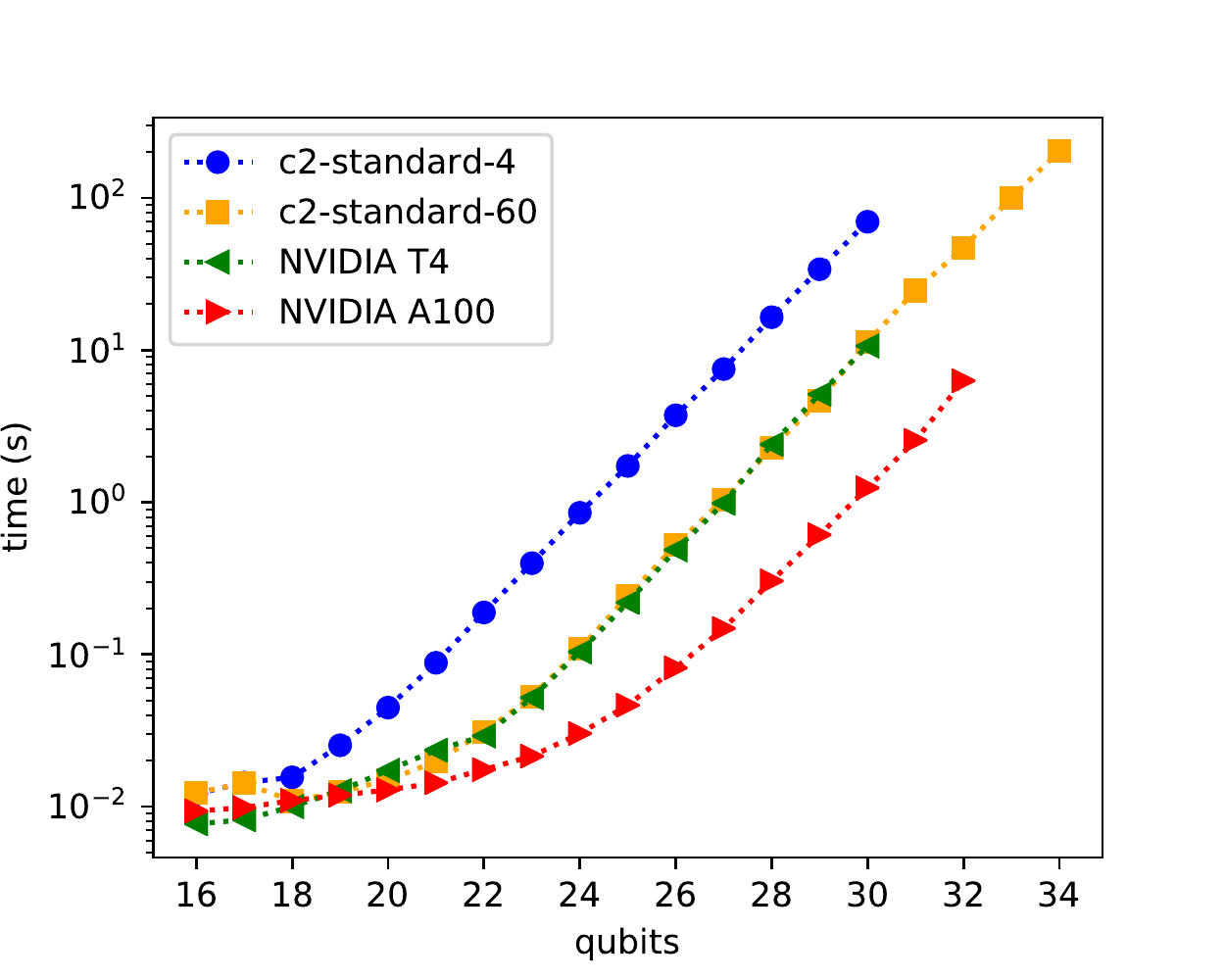}
    \caption{The qsim runtime versus the number of qubits is exponential for large circuits.  We used random circuits with depth 20. The different labels correspond to different machines in Google Cloud Platform. The maximum fuse size is set to 4 and the number of threads is equal to the number of cores in the machine. The speedup between the different machines is roughly a factor of 7.
}
    \label{fig:qubits_vs_time}
\end{figure}

The main factor in the runtime of a circuit simulation is the number of qubits. The size of the state vector for $n$ qubits is $2^n$, and therefore the runtime is also exponential in the number of qubits (for large circuits), as seen in Fig.~\ref{fig:qubits_vs_time}. For best performance, we set the number of threads equal to the number of cores in the machine. If the maximum number of threads is not used on multi-socket machines, then it is advisable to distribute threads evenly to all sockets or to run all threads within a single socket.  Separate simulations on each socket can be run simultaneously in the latter case.  Note that, due to OpenMP overhead, the number of CPU threads does not affect the performance for circuits smaller than 17 qubits (when is better to use one thread). The runtime is linear in the circuit depth, as the number of matrix-vector multiplications is linear in the depth. 

\begin{figure*}[htb]

\subfloat[Time versus noise, 20 qubits.\label{subfig:perf_noise_q20}]{
  \includegraphics[width=0.48\textwidth]{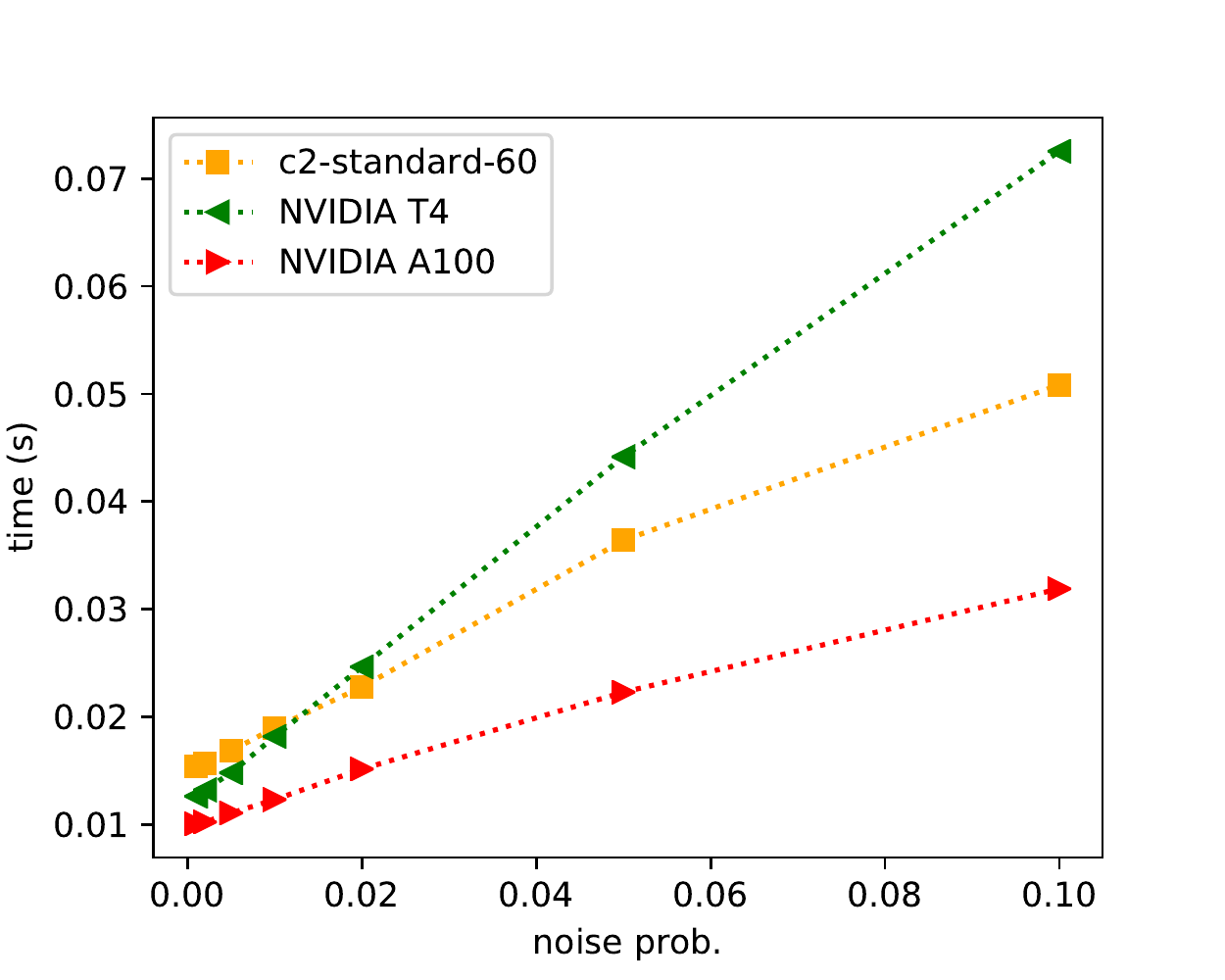}
}\hfill
\subfloat[Time versus noise, 27 qubits.\label{subfig:perf_noise_q27}]{
  \includegraphics[width=0.48\textwidth]{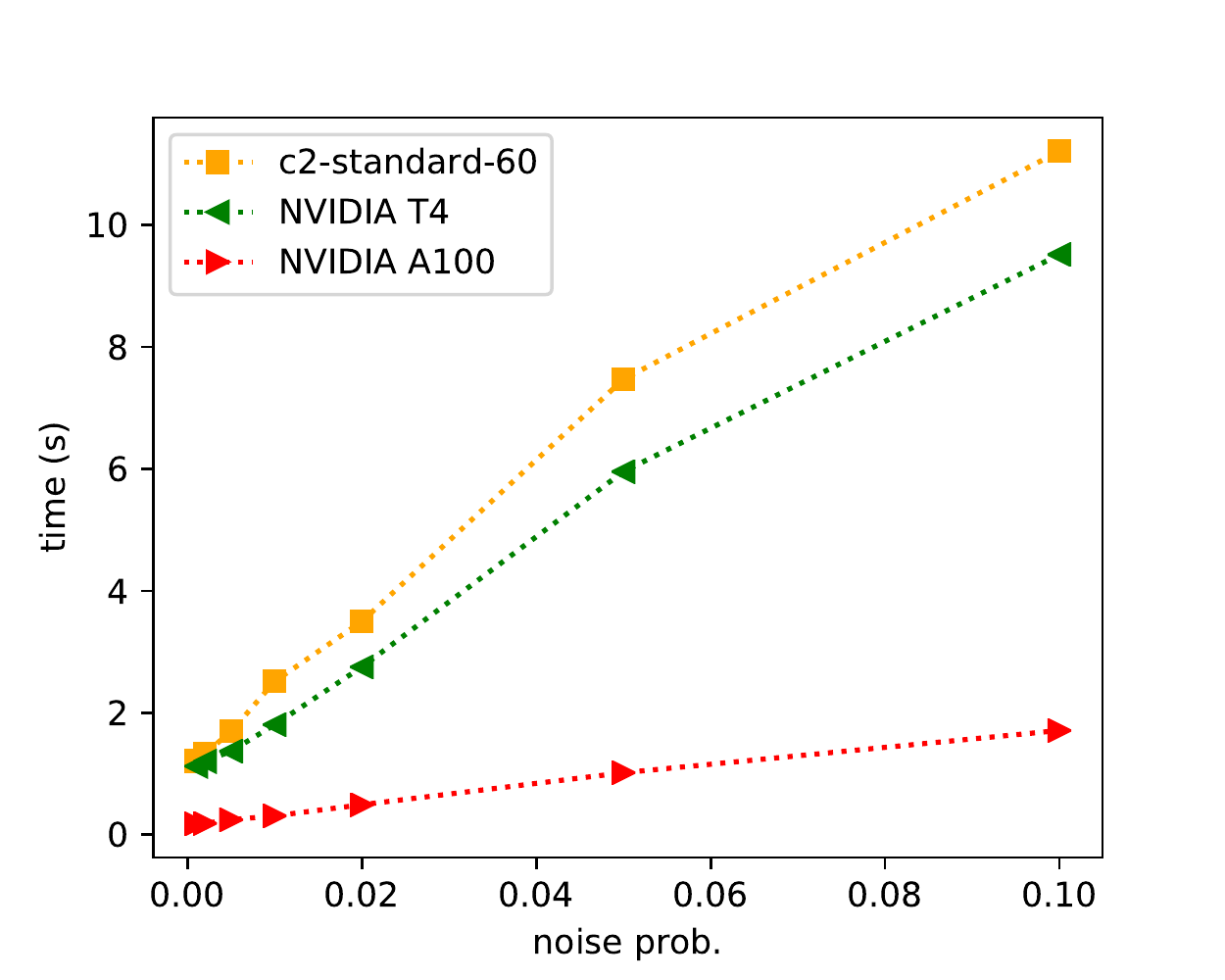}
}

\caption{The qsim runtime (for a quantum trajectory) increases linearly with the noise strength for a phase damping channel. The maximum fuse size was set to 4. Different lines correspond to different machines in Google cloud. }\label{fig:perf_noise}

\end{figure*}

Figure~\ref{fig:perf_noise} shows that the runtime increases linearly with the noise strength for a quantum trajectory.  In this example, we used a phase damping channel applied to random circuits on 20 qubits (Fig.~\ref{subfig:perf_noise_q20}) and 27 qubits (Fig.~\ref{subfig:perf_noise_q27}).  For 20 qubit circuits, the c2-standard-60 CPU outperforms an NVIDIA T4 GPU as noise increases, likely because the whole wave function fits in CPU cache.  For 24 or more qubits, the T4 GPU outperforms the c2-standard-60 CPU, likely because the CPU runtime is limited by RAM access time as the wave function does not fit in cache.  Furthermore, high noise prevents gate fusion and limits the arithmetic intensity.  The performance does not depend on the noise strength if all Kraus operators in the quantum channel are proportional to unitary matrices (like in the depolarizing channel), see Sec.~\ref{sec:trajectory_sim_section}.

\section{Simulating quantum circuits on the Google Cloud Platform}
\subsection{Choosing hardware}\label{sec:qsim_hardware}
The first thing to take into consideration when choosing hardware for a qsim simulation, is the memory required to simulate the circuit. The memory required to simulate an $n$ qubit circuit is $8 \cdot 2^n$ bytes. The maximum number of qubits that can be simulated on a given machine is limited by its RAM memory. Currently, this maximum is 32 qubits in a Google cloud GPU (on an NVIDIA A100 GPU with 40GB of memory), and 40 qubits on a virtual machine (on an m2-ultramem-416), see Sec.~\ref{sec:high_memory}.  Note that memory bandwidth affects qsim performance. qsim performs best when it can use the maximum number of threads, and multi-threaded simulation benefits from high-bandwidth memory. 

Simulations of noise can require more memory (by a factor of two). Extra memory is not required for noise models that can be represented with Kraus operators proportional to unitary matrices (see section \ref{sec:trajectory_sim_section}). An example is a noise model which contains only depolarizing channels.

For circuits that contain fewer than 20 qubits, the qsimcirq translation layer performance overhead currently tends to dominate the runtime of the simulation. In addition to this, qsim is currently not optimized for small circuits. 

Performance is the key differentiator between CPUs and GPUs. GPU hardware outperforms CPU hardware significantly (see Fig.~\ref{fig:qubits_vs_time}) when the simulated circuit contains more than 20 qubits. Performance is particularly important in a noise simulation with many trajectories. The run time of a quantum trajectory simulation increases linearly with the number of trajectories. Another case where performance is particularly important is when simulating parametrized circuits for many different choices of parameters. These simulations are embarrassingly parallelizable and well suited for multinode simulations in Google Cloud Platform, see Sec.~\ref{sec:multinode}.

\subsection{Simulation on a high memory CPU}\label{sec:high_memory}

qsim simulations on Google Cloud Platform (GCP) take place in virtual machines (VMs). A VM on GCP behaves like a real computer with user-defined specifications, such as CPU type and available memory, but without being tied to a specific physical device. This allows VMs to take advantage of available resources fluidly, without requiring the user to manage hardware details. GCP VMs also provide users access to more powerful devices than most desktop machines; these include high-memory devices which can support up to 40 qubits in noiseless simulations.

A single GCP VM is sufficient for noiseless qsim simulations, or for qsim quantum trajectory simulations of up to around 23 qubits. A step-by-step workflow for configuring such a simulation on either CPU or GPU is provided on the Quantum AI website.\footnote{\href{https://quantumai.google/qsim/tutorials/gcp\_cpu}{https://quantumai.google/qsim/tutorials/gcp\_cpu}} The outline of the CPU workflow is as follows:

\begin{enumerate}
\item Create a VM with the necessary specifications for the simulation, see Sec.~\ref{sec:qsim_hardware}.
\item Establish an ssh connection from your computer to the VM.
\item Start the provided qsim Docker container on the VM. The qsim Docker container is a self-contained environment with qsim and its dependencies already installed. 
\item Run qsim simulations on the VM. The three main options for how to run qsim simulations: via Google Colab,\footnote{\href{https://colab.research.google.com/}{https://colab.research.google.com/}} Jupyter,\footnote{\href{https://jupyter.org/}{https://jupyter.org/}} or directly in the terminal.
\end{enumerate}

\subsection{Multinode simulation}\label{sec:multinode}

\begin{figure*}
    \centering
    \includegraphics[width=\linewidth]{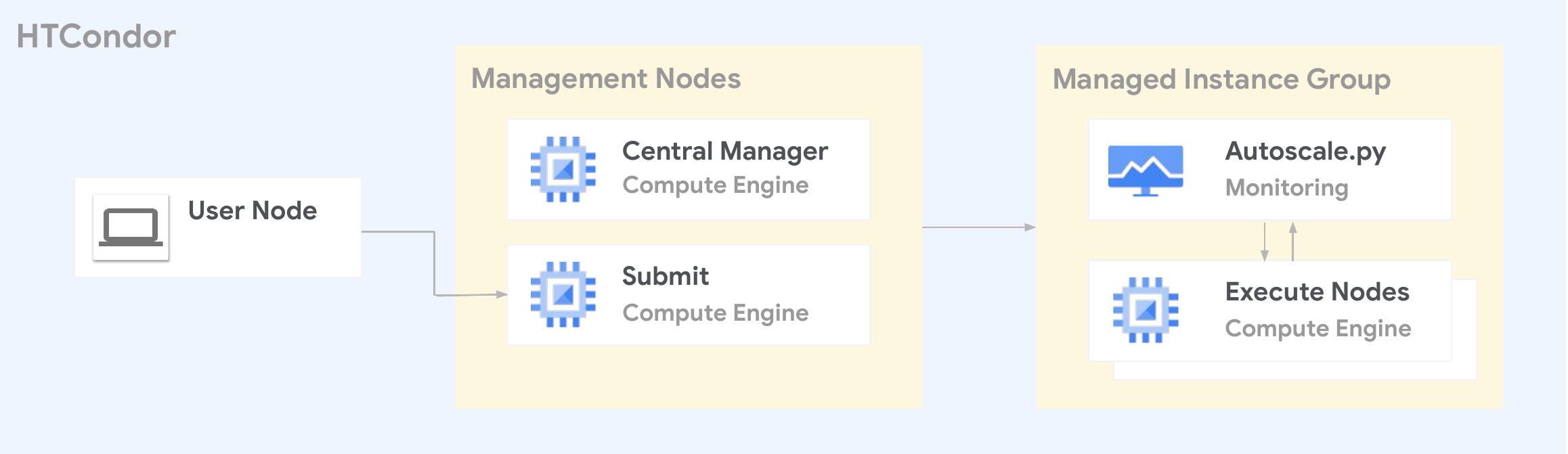}
    \caption{HTCondor workflow for simulations on multiple compute nodes}
    \label{fig:HTcondor}
\end{figure*}

Quantum trajectories of noisy quantum circuits are embarrassingly parallelizable. Therefore a simulation can be distributed over multiple compute nodes to finish it linearly faster in the number of nodes. One way to do this in GCP is using TensorFlow Quantum and Kubernetes with qsim as the simulation backend~\cite{xing_blog_2021}. This is particularly natural for research in quantum machine learning applications.  
In this section, we explain a different workflow for configuring multinode simulations on GCP using a more traditional job scheduler. We provide a complete tutorial on the Google Quantum AI  website.\footnote{\href{https://quantumai.google/qsim/tutorials/multinode}{https://quantumai.google/qsim/tutorials/multinode}} The scripts initialize a cluster and include a job submit file that can be modified to call a custom Python script.

The multinode simulation workflow creates the virtual machines required for the simulation using autoscaling. The autoscaler is able to cope with heterogeneous clusters (containing multiple machine types in multiple regions), using HTCondor matchmaking algorithms to require or prefer co-locating compute and storage for optimal performance.

The workflow provides Terraform\cite{terraform} scripts to create the HTCondor\cite{htcondor} cluster. Terraform is an Infrastructure as Code platform which allows cloud infrastructure to be created, managed and destroyed using a well defined programming model. When called, the Terraform code interacts with the Google Cloud APIs to create several components (also known as \emph{resources}) of the cluster, see Fig.~\ref{fig:HTcondor}:
\begin{enumerate}
    \item A \emph{Managed Instance Group} (MIG), which represents a group of Virtual Machines (VM) based on a template
    \item An \emph{Instance Template}, which provides the definition of the VMs used in the MIG
    \item A \emph{controller node} (or Central Manager) which manages all aspects of the cluster
    \item A \emph{submit node}, which allows users to login and interact with the cluster
\end{enumerate}

The controller node runs a Negotiator and a Collector process. These processes maintain lists of all the nodes that are part of the HTCondor cluster (also referred to as a \emph{pool}). All HTCondor compute nodes are registered by the controller. Compute nodes can be added or removed from the Collector. In the cluster created by the Terraform in the provided repository, the default configuration does not create compute nodes. Only when the work is detected in the submit queue are compute nodes created.  This is the function of the autoscaler.

The autoscaler runs on a 60 second schedule, detecting if jobs are idle in the submit queue, due to a lack of compute resources. If a lack of resources are detected, the autoscaler will send a signal to the MIG to increase the number of nodes. The autoscaler will request new resources in this way until the defined limit is reached (defaults to 20 nodes, which can be changed as needed). The MIG responds by creating new VMs (compute nodes) from the Instance Template created by the Terraform. The template includes the HTCondor system processes for the compute node. These processes connect to the Central Manager and are subsequently registered with the Collector process.

The Negotiator process is now responsible to connect the idle jobs in the queue with the newly created and registered resources created by the MIG. The Negotiator makes use of matching algorithms to determine which jobs should be run on which compute nodes. In the default case, all compute nodes are created from the same template, and thus have homogeneous configurations, but in the case of heterogeneous configurations, the Negotiator will assign jobs to nodes with appropriate resources.

Finally, jobs are executed on their assigned compute nodes. When completed, the job results are (in the default case) returned to the \emph{out} directory on the submit node. At this point, the job is removed from the queue and a waiting idle job may take over the resources of the now idle compute node. 

At the end of the multinode simulation, it is a simple matter to destroy the HTCondor cluster with all assigned resources, using a \emph{terraform destroy} command.

\subsection{A multinode example: $\mathbb{Z}_2$ gauge
theory quantum circuits}

\begin{figure}
    \centering
    \includegraphics[width=\columnwidth]{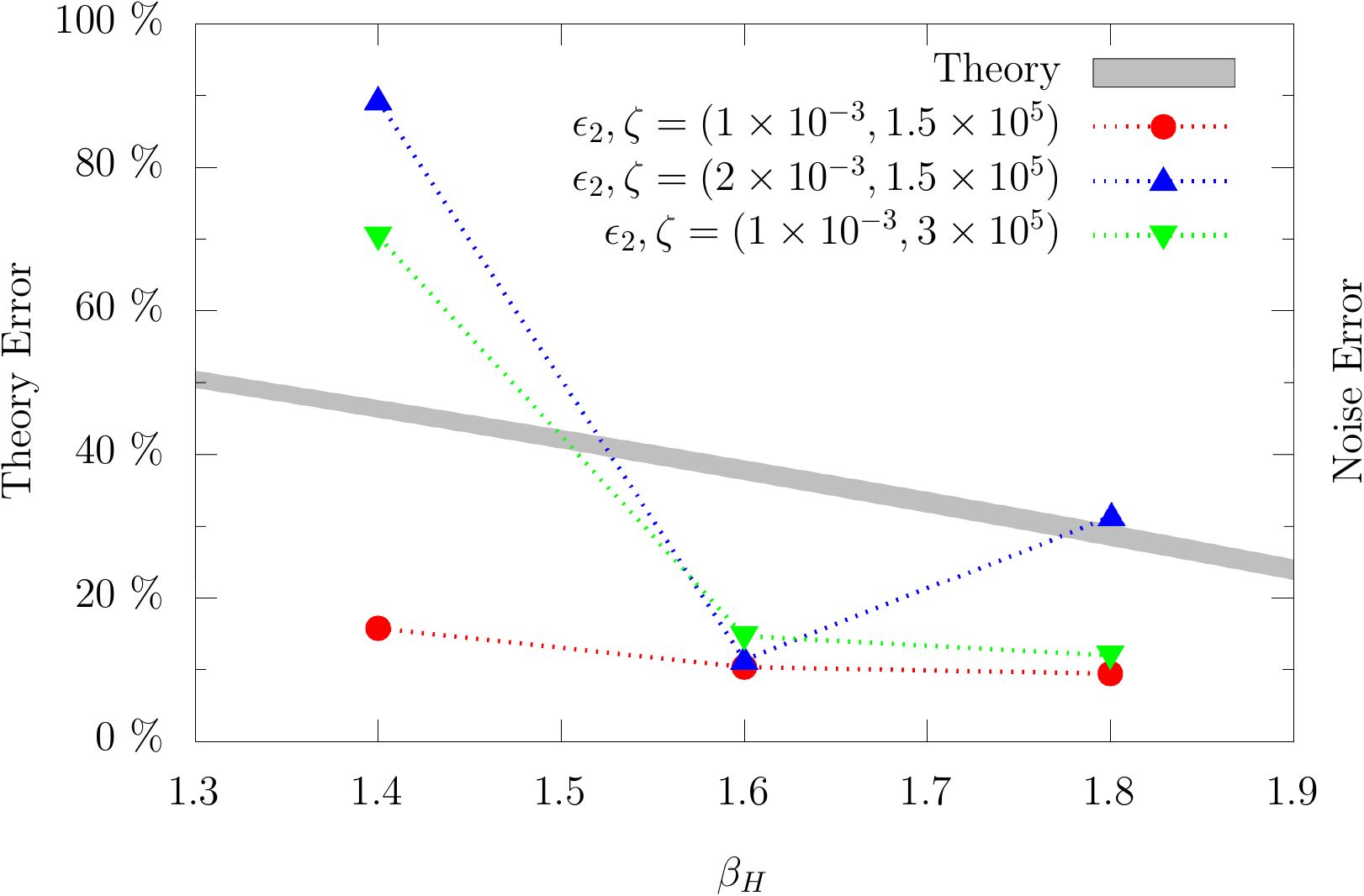}
    \caption{Sources of relative systematic error in computed mass as a function of $\beta_H$, which for increasing value decreases errors lattice errors (at the cost of finite volume errors), see Ref.~\onlinecite{gustafson2021large}. 
    The gray band indicates the estimated theoretical errors from extrapolating with $N=3,4,5,6$ classical lattices.
    The error from noise are shown for different fiducial noise models.
    $\epsilon_2$ is the two-qubit depolarizing noise parameter, and $\zeta$ is the control error scaling parameter.}
    \label{fig:err_comp}
\end{figure}

Obtaining predictions from quantum field theory (QFT) requires overcoming a doubly infinite-dimensional problem to use finite computational resources.  The first is that QFTs are defined in the infinite, continuous volume. This obstacle is overcome by computing the theory on a lattice and extrapolating to the continuum. The second infinity is the quantum fields at each point can take any value and thus must be truncated.
Limitations in the effective qubit connectivity of near-term quantum computers constrain the choice of models.
In Ref.~\onlinecite{gustafson2021large}, we chose to simulate the $\mathbb{Z}_2$ gauge theory -- a simple lattice theory with 1 qubit per link on a two-dimensional lattice with length $N$; this allows for mapping to a square arrangement of $N^2$ qubits.
This theory represents a low rung on a ``ladder'' of theories we can climb to approach nature.

Our goal was to begin to define the boundary for a beyond-classical calculation in lattice field theory.
We chose a property - the lightest mass of a particle - of the theory we \emph{could} calculate classically as a benchmark --- our assumption is uncertainties on this quantity will look similar to uncertainties on other quantities computed on the quantum computer.
We simulated lattices of increasing $N=3,4,5,6$.
A full state vector simulation for a $N^2=7^2$ lattice is infeasible on existing supercomputers while simulation of an $8^2$ lattice is completely out of reach.
By harnessing the  computational capabilities of GCP, and a special TPU implementation, we simulated quantum circuits of 36 qubits to see if it could compute our benchmark.
The resulting computation showed very poorly controlled uncertainties, implying we require at least the $7^2$ lattice, and possibly more (there is no way to be sure without seeing results from the 49 qubit machine).
These circuits are essentially out of reach for exact circuit simulation, although approximate quantum simulation may produce results with acceptable uncertainties.

Another important aspect of this research was to understand how noise on a quantum computer impacts our results.
Currently available quantum computers cannot execute the large circuits required for QFT - typically tens of thousands of single and two-qubit gate operations -- with high fidelity. To better understand the feasibility of QFT simulations on the next generation of improved NISQ devices we studied a simple noise model based on local, uncorrelated errors. For each $n$-qubit gate we implemented a depolarizing channel on the corresponding $n$-qubit subsystem,
\begin{equation}\label{eq:depol_1}
    \mathcal{D}_n[\epsilon](\rho) = (1 - \epsilon) \rho + \frac{\epsilon}{2^n} I
\end{equation}
where $\rho$ is the density matrix, and also modelled control errors for $\sqrt{\text{iSWAP}}$ gates according to the unitary map $\Lambda_{ZZ}( \rho) = U_{ZZ} \rho U_{ZZ}^\dagger$ on each 2-qubit subsystem, with a unitary error term scaled by a parameter $\zeta$ as:
\begin{equation}\label{eq:zeta_noise}
    U_{ZZ}[\zeta] = \exp \left(-i2\pi \zeta T |11\rangle\langle 11| \right)\;,
\end{equation}
where $T$ is the gate time (roughly 10 ns). 
Under this assumed model for device noise, we found that hardware errors dominate the theory error arising from the discretization of the QFT (Fig. \ref{fig:err_comp}). Improvements of this magnitude will require significant engineering breakthroughs, but as quantum error correction becomes more widely available, studying QFT problems of scientific interest may become possible, perhaps without even requiring fully fault-tolerant devices.

\section{Approximate noise simulation}

\subsection{Main QCS error measurements} \label{errors_overview}

We describe here some of the main error measurements in Google's Quantum Computing Service (QCS).  Although not comprehensive, they capture the dominant error mechanisms for superconducting qubits. 

\subsubsection{Qubit decay and dephasing}

Incoherent errors represents error that cannot be compensated by improved qubit control or unitary operations. Typically they are associated with interactions with an external environment. The first incoherent error mechanism is decay of the qubit from the excited state $\ket 1$ to the ground state $\ket 0$. This process is characterized by preparing the excited state with a microwave pulse (a.k.a. an X gate), then measuring the survival probability as a function of time $t$. This probability is fit to an exponential decay $e^{-t/T_1}$ to determine the characteristic decay time $T_1$ (reported in microseconds).

The second incoherent error mechanism is dephasing. Unlike decay, it does not cause transitions between the qubit's energy states. Rather it causes coherent superpositions of $\ket 0$ and $\ket 1$ to ``collapse'' into one state or the other. Typically dephasing is characterized by a dephasing rate $1/T_{\phi}$, usually extracted through Ramsey or spin echo experiments. The dephasing rates extracted by these experiments vary since they probe the noise at different frequency scales. For simplicity we instead choose to infer the dephasing rate through a measurement of the total incoherent error $\epsilon_{inc}$. For one-qubit gates $\epsilon_{inc}$ can be measured using purity benchmarking~\cite{feng2016estimating}. The white noise dephasing time $T_{\phi}$ is estimated to leading order from the incoherent error $\epsilon_{inc}$ as
$$\epsilon_{inc}=\frac{t}{3T_1} + \frac{t}{3T_\phi}+O(t^2)\;.$$

\subsubsection{Parallel readout errors}

Readout errors correspond to a qubit being measured in computational state $\ket {{\rm not} b}$ with $b \in \{0,1\}$ while actually being in state $\ket {b}$. The readout error probabilities are measured in parallel for all qubits in parallel, to account for readout crosstalk. These values are measured by preparing all qubits in random computational basis states, then reading their value. The value \lstinline{parallel_p00_error} is estimated as the fraction of the time that state $\ket{0}$ is measured as $\ket{1}$. The value \lstinline{parallel_p11_error} is estimated as the fraction of the time that state $\ket{1}$ is measured as $\ket{0}$

Readout error is affected by several error mechanisms.
First, the signal used to infer the qubit state is subject to classical noise, which can cause the qubit state to be misidentified. Additionally, the qubit may decay from $\ket 1$ to $\ket 0$ during the measurement process. The probability \lstinline{parallel_p11_error} is hence generally expected to be higher than \lstinline{parallel_p00_error}. In addition to this, there is potential for additional errors caused by unintended readout line crosstalk or interactions with other qubits.

\subsubsection{Isolated one-qubit gate RB error}

Average one-qubit gate error is estimated by a technique known as randomized benchmarking (RB)~\cite{magesan2011scalable,magesan2012characterizing}. This is done by applying gate sequences of varying length composed of randomly chosen one-qubit Clifford operations (i.e. the group of unitaries  that preserve the Pauli group under conjugation). The final operation in each sequence is chosen to invert the product of all predecessors, so the result of the total sequence should always be the identity. The success probability (i.e. probability of measuring the initial state $\ket 0$) is averaged over many sequences of increasing lenghts and is fitted to an exponential decay. The average one-qubit gate error is extracted from this fit. This isolated error is calculated for one qubit at a time while all other qubits on the device are idle. 

\subsubsection{Two-qubit gate parallel XEB error}

Two-qubit gate error is primarily characterized by applying cross-entropy benchmarking (XEB)~\cite{boixo2018characterizing,neill2018blueprint,arute2019quantum}. This procedure repeatedly performs a ``cycle'' of a random one-qubit gate on each qubit followed by the two-qubit entangling gate. The resulting distribution is analyzed and compared to the expected distribution using cross entropy. This is averaged over many sequences of increasing lenghts, as in RB. The value reported is the error rate of multiple parallel 2-qubit cycles at a time. Four different discrete patterns of 2-qubits are used, with each pair of qubits in only one pattern.

Since there are many different possible layouts of parallel two-qubit gates and each layout may have different cross-talk effects, users may want to perform this experiment on their own if they have a specific layout commonly used in their experiment.

\subsection{Approximate noise model}\label{sec:approx_noise}

To approximate the QCS hardware noise in a quantum circuit simulation, we add four types of noise to the circuit: decay and dephasing errors, readout errors, entangling gate errors and depolarizing errors. The noise channels corresponding to these noise categories consume processor specific calibration data (see Sec.~\ref{errors_overview}).

\subsubsection{Decay and dephasing errors}

A decay rate $1/T_1$ and a pure dephasing rate $1/T_{\phi}$, both occurring over time $t$, result in the quantum channel~\cite{nielsen2002quantum} 
$$\mathcal{E}(\rho)=\twoMat{1-\rho_{11} e^{-t/T_1}}{\rho_{01}e^{-t/T_2}}{\rho_{10}e^{-t/T_2}}{\rho_{11}e^{-t/T_1}}$$
where
$$\frac{1}{T_2}=\frac{1}{2T_1} + \frac{1}{T_{\phi}}$$
The channel $\mathcal{E}$ can be written using three Kraus operators,
\begin{align}
    \begin{split}
        K_0 & = \twoMat{1}{0}{0}{e^{-t/T_2}}\\
        K_1 & = \twoMat{0}{\sqrt{1-e^{-t/T_1}}}{0}{0}\\
        K_2 & = \twoMat{1}{0}{0}{\sqrt{e^{-t/T_1}-e^{-2t/T_2}}}
    \end{split}
\end{align}
The channel $\mathcal{E}$ is applied on each qubit after every gate or idle time. Sample data for gate durations and qubit decay times can be found in the QCS datasheet\footnote{https://quantumai.google/hardware/datasheet/weber.pdf}, in \cite{2020} and in \cite{2021}.

\subsubsection{fSim gate coherent errors}

As described in Ref.~\onlinecite{2020}, a Fermionic Simulation or fSim gate can be represented as the following matrix
\begin{equation}
    {\rm fSim}(\theta,\phi)= \begin{pmatrix} 
        1 & 0 & 0 & 0 \\
        0 & \cos(\theta) & -i \sin(\theta)  & 0 \\
        0 & -i \sin(\theta)  & \cos(\theta) & 0 \\
        0 & 0 & 0 & e^{-i\phi}
        \end{pmatrix}
\end{equation}
Where $\theta$ represents the $\ket{01}\leftrightarrow \ket{10}$ SWAP angle and the phase $\phi$ on state $\ket{11}$ is the CPhase angle. Particular choices of fSim angles produce a variety of gates such as CZ, iSWAP, and $\sqrt{\text{iSWAP}}$.

fSim angles are optimized to maximize the parallel cross entropy benchmarking (XEB) fidelity. To model the deviations $\delta \theta$ and $\delta \phi$ from the desired angles, we simply apply the gate fSim$(\delta \theta, \delta \phi)$ after each fSim gate. Additionally, we also include single qubit $Z$ phase errors $e^{i\varphi Z}$ before and after the fSim gate. The $Z$ phase angle errors can be obtained by sampling a typical distribution obtained with Floquet calibration (see Ref.~\onlinecite{arute2020observation}).

\subsubsection{Depolarizing errors}

We use the depolarizing channel to account for any additional error not explicitly included above. The amount of depolarizing error added is chosen so that the total Pauli error matches values measured in either one-qubit RB or two-qubit XEB. See Ref.~\onlinecite{2019} for formulas on the conversion between different error rate standards. 
The depolarizing error is represented using the standard depolarizing channel
\begin{equation}
    \mathcal{E}_{dep}(\rho)= (1-r_{\rm dep})\rho + \frac{r_{\rm dep}}{D^2-1}\sum_{\mu\neq 0} P_\mu \rho P_\mu
\end{equation}
where $D$ is the dimension of the system (2 or 4, depending on qubit number) and the sum is taken over all Pauli operators excluding the identity. 

For-two qubit gates, we infer the depolarizing Pauli error $r_{\rm dep}$ from the total XEB Pauli error $r_p^{\rm tot}$ by subtracting the incoherent error $r_{\rm inc}$ (on each qubit) and the entangling error $r_{\rm ent}$ (from the coherent errors on the fSim gate) 
\begin{equation}
    r_p^{\rm tot} = r_{\rm inc}^0 + r_{\rm inc}^1 + r_{\rm ent} + r_{\rm dep}\;.
\end{equation}

\subsection{Examples: simulating experiments on GCP with approximate noise}

We compare experimental implementations in QCS of several quantum algorithms with numerical simulations using the approximate noise model detailed in Sec.~\ref{sec:approx_noise}. We also do numerical simulations using a simpler noise model consisting of a depolarizing channel after each gate. The depolarization strength is set to give the RB fidelity for one-qubit gates or the XEB fidelity for two-qubit gates (see above). We see that the approximate noise model is closer to the experimental data than the simpler depolarizing noise.

\subsubsection{Fermi-Hubbard interacting dynamics}

We compare the dynamical evolution of four interacting fermions as described in Ref.~\onlinecite{arute2020observation} between a experimental GCP implementation and simulations. The specific noise parameters for each one-qubit and each two-qubit gate are obtained from the device's characterization performed at the same day of the experiment.

\begin{figure}
    \centering
    \includegraphics[width=\linewidth]{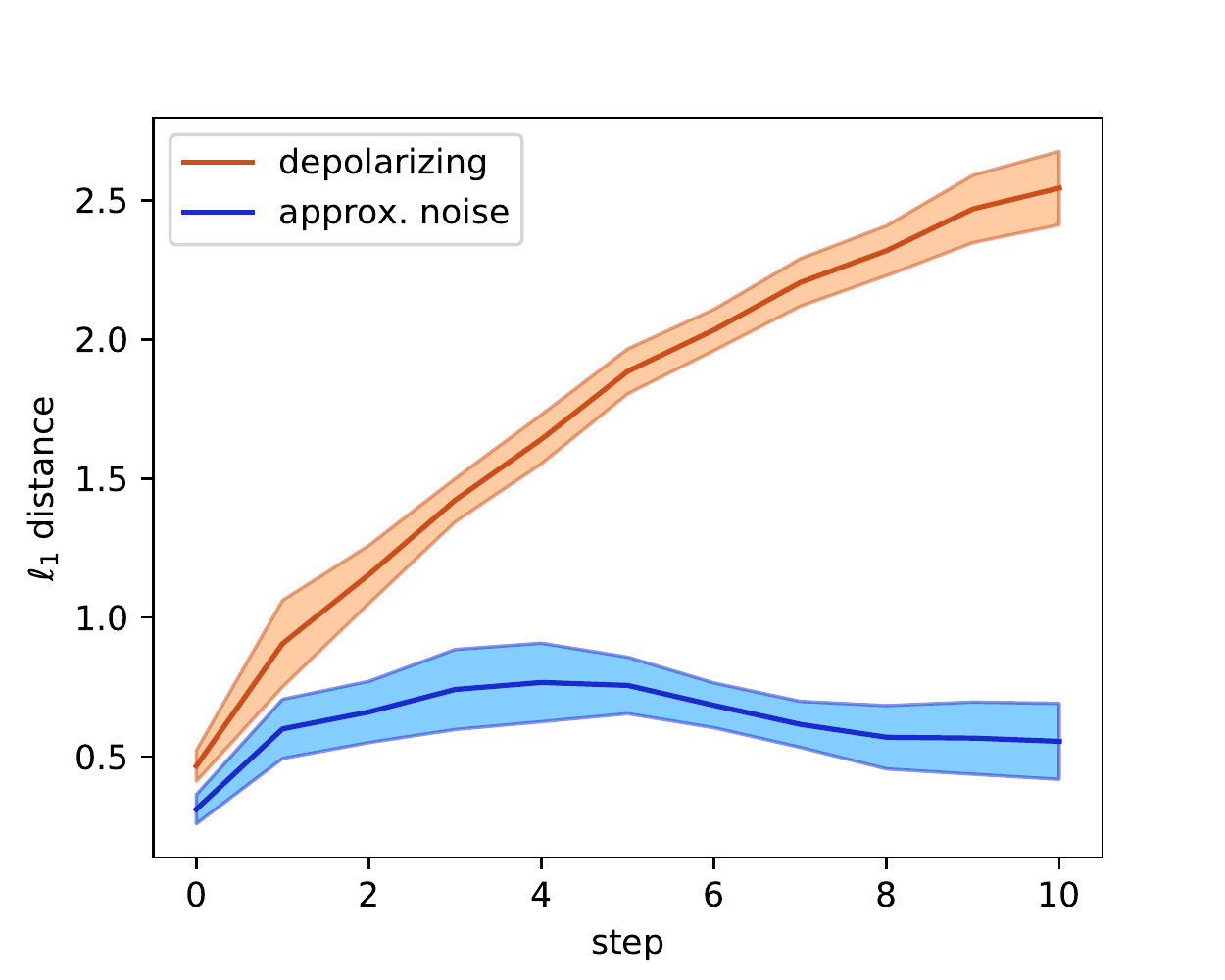}
    \caption{Fermi Hubbard model dynamics evolution experiment in Ref.~\onlinecite{arute2020observation} on a quantum experimental processor compared against an approximate noise simulation. The instance used for this benchmark is the 4 fermionic interacting instance from Ref.~\onlinecite{arute2020observation} with 8 spin-up and spin down sites, 2 fermions at each site and interaction strength $u = 4$. The distance between the experimental device and noisy simulation is calculated as the $\ell_1$ distance between fermionic distributions at each site. Multiple experiments were performed on 16 different qubit configurations leading to the standard deviation shown.
}
    \label{fig:FH_plot}
\end{figure}

Figure \ref{fig:FH_plot} shows the result of this comparison. At each Trotter step, we measure the fermionic particle densities, that is the probability of measuring the $Z$ up state $\ket 1$ at each site. We repeat the experiment on both the quantum device and the simulator with approximate noise for 16 different qubit arrangements and calculate the $\ell_1$ distance 
\begin{align}
    \ell_1 = \sum_j |q_j - p_j|\;,
\end{align}
where $q_j$ is the experimental fermionic density distribution (probability of $\ket 1$), $p_j$ is the numerically simulated one, and the sum is over all qubits or sites. Each qubit arrangement is characterized by different noise parameters, which leads to a distribution of measured distances. 

Contrary to the experiments in Ref.~\onlinecite{arute2020observation}, the experiments used for this comparison do not use post-selected and re-scaled results. Floquet calibration was used on both device experiment and simulation. The results are averaged over many qubit configurations after calculating the $\ell_1$ distance between them.

\subsubsection{QAOA}

\begin{figure}
    \centering
    \includegraphics[width=\linewidth]{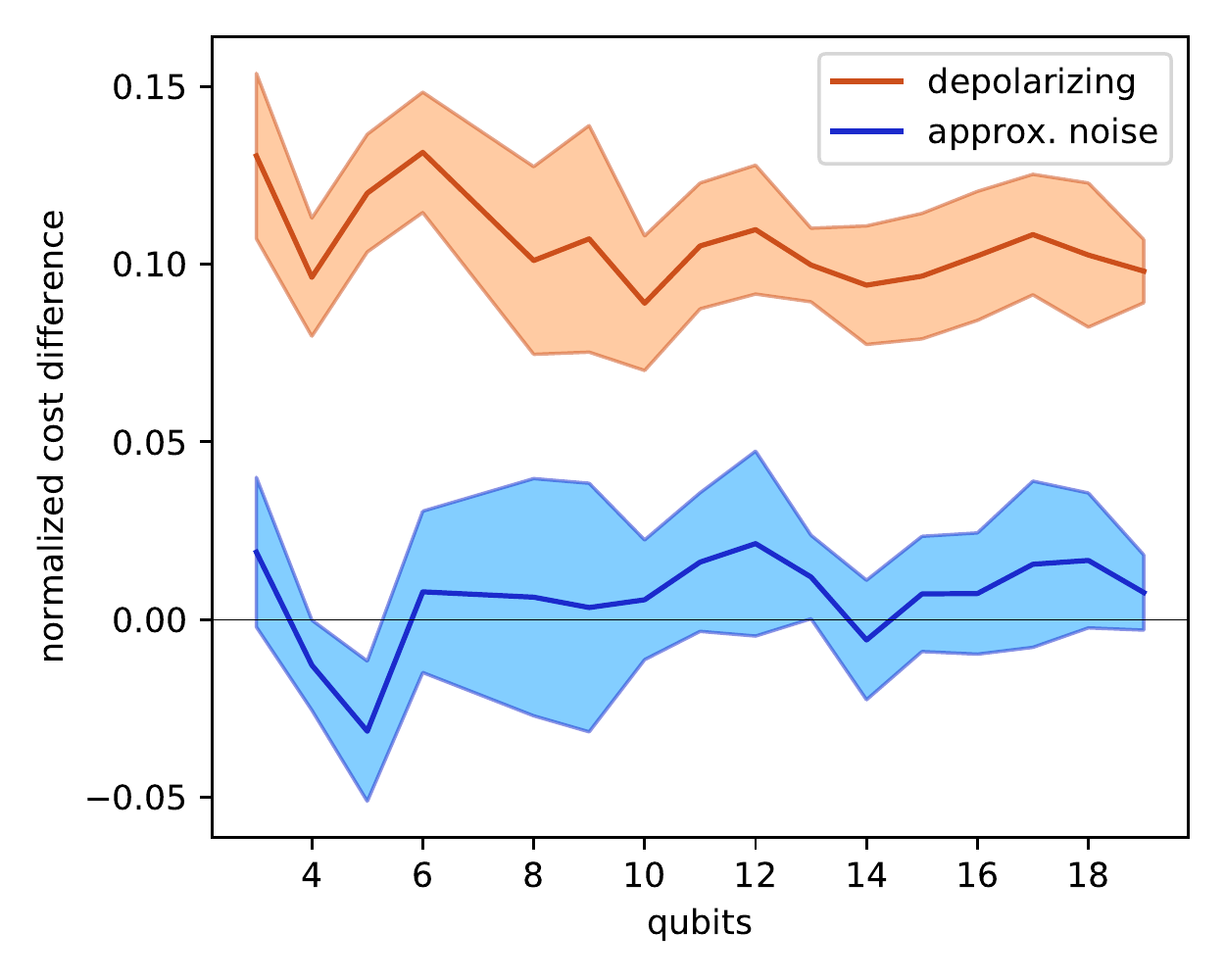}
    \caption{Difference between quantum hardware and noise simulation with QAOA on ``Hardware Grid" problems over a variety of sizes with depth $p=3$.}
    \label{fig:QAOA_plot}
\end{figure}

Figure~\ref{fig:QAOA_plot} shows a comparison between a GCP experimental implemenation and simulations for the quantum approximate optimization algorithm (QAOA).  This algorithm aims to solve combinatorial optimization problems with a heuristic application of Z-basis entangling operations and X-basis mixing operations. Following the experiment described in Ref.~\onlinecite{harrigan2021}, we measure its performance using the average fraction of satisfied clauses relative to the global optimum: $\langle C \rangle / C_\mathrm{min}$. In Fig.~\ref{fig:QAOA_plot}, we plot the difference in this metric between noise simulations and hardware results for ``Hardware Grid" problems \cite{harrigan2021} of varying problem sizes (i.e. qubit number). The QAOA depth hyperparameter $p$ is set to three. The approximate noise model results closely match those from the hardware, while a simple depolarizing noise model is overly optimistic.

\subsubsection{Non-interacting fermion dynamics}

\begin{figure}
    \centering
    \includegraphics[width=8.5cm]{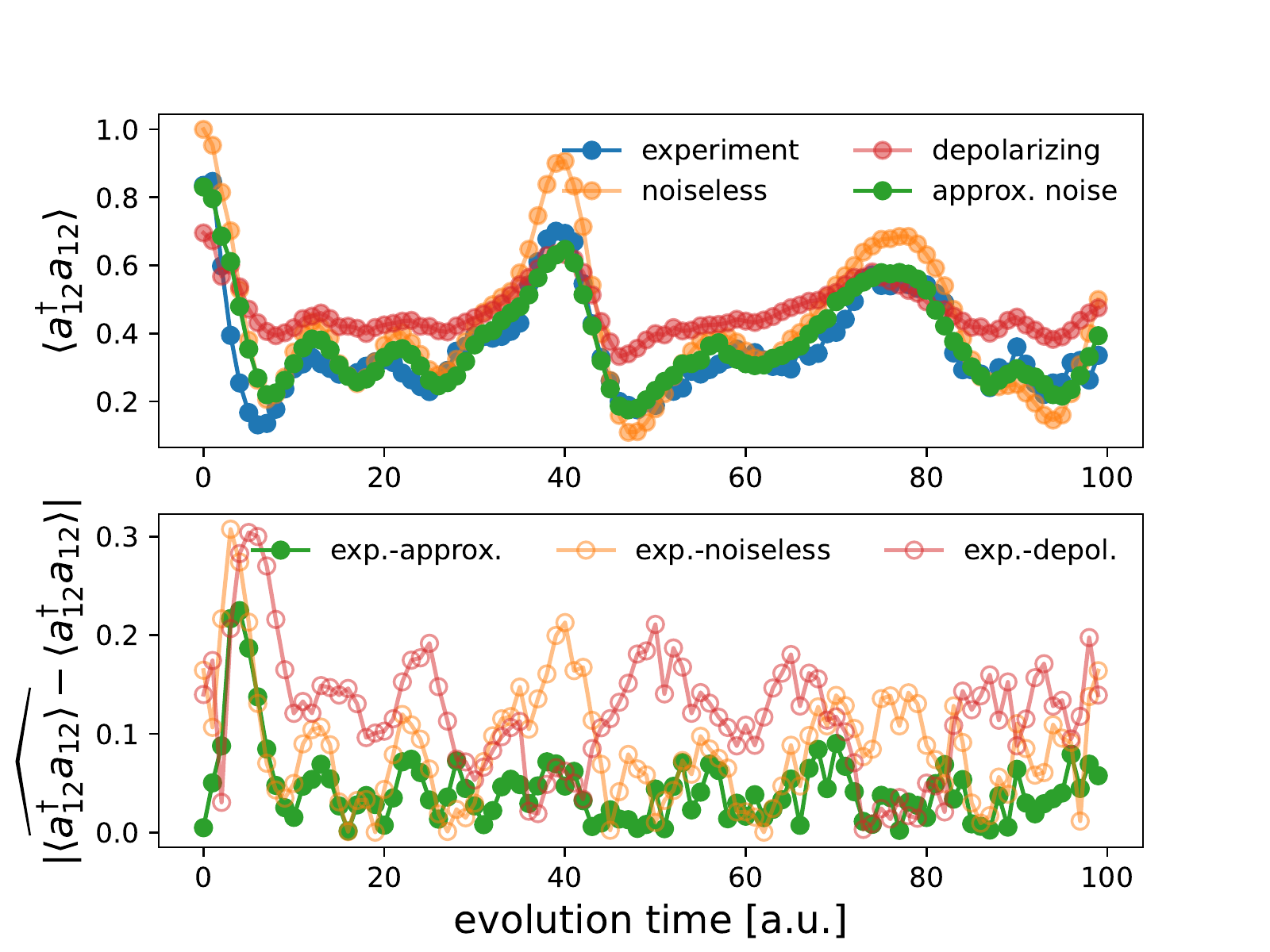}
    \caption{Top: Local occupation of the twelfth qubit in a linear array implementing unitary dynamics of $H_{\rm n.n.}$ for time $t \in [0, 20)$, in arbitrary units, with (blue) the Sycamore device (QC), (orange) noiseless simulation, (red) noise simulation with a 1\% 1-qubit depolarizing channel applied after each layer of qubits, and (green) the approximate noise model.  Bottom: Absolute difference between the experimental QCS data (QC) and each corresponding simulation method (see top panel).  Lower values for the approximate noise model indicates a better approximate to the experimental device. }
    \label{fig:ncr_dynamics}
\end{figure}

As a third example, we implement non-interacting fermion dynamics on a line of qubits. These circuits are the basis of many fermionic simulation algorithms and compilation strategies~\cite{PhysRevLett.120.110501, Rubinscience.abb9811, PhysRevApplied.9.044036, rubin2021compressing}.  We start with a simpler problem of evolving a fermionic system with a nearest-neighbor hopping Hamiltonian:
\begin{align}
H_{\rm n.n.} = -1 \sum_{i=1}^{L-1}\left(a_{i}^{\dagger}a_{i+1} + a_{i+1}^{\dagger}a_{i}\right)
\end{align}
where $L$ is the total number of qubits. The unitary $U(t)$ that is generated by $e^{-i t H_{\rm n.n.}}$ can always be implemented in linear depth and quadratic number of gates for arbitrary time $t$~\cite{jozsa2008matchgates, clements2016optimal, PhysRevLett.73.58}.  Specifically, each $U(t)$ is decomposed into a sequence of nearest-neighbor two-qubit gates called Givens rotation gates, which can in turn be synthesized with two $\sqrt{\mathrm{iSWAP}}$ gates and two Z-rotations each.  The Givens rotation synthesis of $U(t)$ will always result in a quadratic number of two-qubit gates regardless of the magnitude of $t$.  This is an instance of so-called "fast-forwardable" dynamics.

We evolved a five electron wave function in a thirteen orbital product state using $U(t)$ for $t \in [0, 20)$ with time steps of $\delta t = 1/5$ (arbitrary units).  The initial product state we used was a determinant corresponding to a computational basis state $\vert 1001001001001\rangle$.  After each computation, all qubits were measured.  Along with running these computations on the device, we also simulated them with the approximate noise model, the ideal unitary, and a simple noise model with one-qubit 1\% depolarizing channel applied to all qubits after each time slice of the circuit.  In Fig. ~\ref{fig:ncr_dynamics}, we plot the expectation value of finding the qubit in the $\vert 1\rangle$ state for the last qubit, and the difference with the simulations.  It is apparent that the approximate noise model provides a better representation of the device than the generic depolarizing noise model.


\section{Closing remarks}
We have provided, as part of Google's Quantum AI open source ecosystem, a high performance quantum trajectory simulator. Quantum trajectory simulations are embarrassingly parallelizable, and therefore well suited for a GCP multinode workflow. We provide the necessary tools to manage an autoscaling pool of VMs with HTCondor. Finally, we explain the dominant errors of superconducting qubits, and provide an approximate error model for Google's QCS. We compare experimental implementations in QCS and approximate simulations for three quantum algorithms. 

Simulation with approximate noise is expected to advance NISQ applications research, by allowing researchers to account for the dominant hardware error mechanisms in real NISQ devices when prototyping algorithms.

\bibliography{main}

\begin{thebibliography}{37}%
\makeatletter
\providecommand \@ifxundefined [1]{%
 \@ifx{#1\undefined}
}%
\providecommand \@ifnum [1]{%
 \ifnum #1\expandafter \@firstoftwo
 \else \expandafter \@secondoftwo
 \fi
}%
\providecommand \@ifx [1]{%
 \ifx #1\expandafter \@firstoftwo
 \else \expandafter \@secondoftwo
 \fi
}%
\providecommand \natexlab [1]{#1}%
\providecommand \enquote  [1]{``#1''}%
\providecommand \bibnamefont  [1]{#1}%
\providecommand \bibfnamefont [1]{#1}%
\providecommand \citenamefont [1]{#1}%
\providecommand \href@noop [0]{\@secondoftwo}%
\providecommand \href [0]{\begingroup \@sanitize@url \@href}%
\providecommand \@href[1]{\@@startlink{#1}\@@href}%
\providecommand \@@href[1]{\endgroup#1\@@endlink}%
\providecommand \@sanitize@url [0]{\catcode `\\12\catcode `\$12\catcode
  `\&12\catcode `\#12\catcode `\^12\catcode `\_12\catcode `\%12\relax}%
\providecommand \@@startlink[1]{}%
\providecommand \@@endlink[0]{}%
\providecommand \url  [0]{\begingroup\@sanitize@url \@url }%
\providecommand \@url [1]{\endgroup\@href {#1}{\urlprefix }}%
\providecommand \urlprefix  [0]{URL }%
\providecommand \Eprint [0]{\href }%
\providecommand \doibase [0]{http://dx.doi.org/}%
\providecommand \selectlanguage [0]{\@gobble}%
\providecommand \bibinfo  [0]{\@secondoftwo}%
\providecommand \bibfield  [0]{\@secondoftwo}%
\providecommand \translation [1]{[#1]}%
\providecommand \BibitemOpen [0]{}%
\providecommand \bibitemStop [0]{}%
\providecommand \bibitemNoStop [0]{.\EOS\space}%
\providecommand \EOS [0]{\spacefactor3000\relax}%
\providecommand \BibitemShut  [1]{\csname bibitem#1\endcsname}%
\let\auto@bib@innerbib\@empty
\bibitem [{\citenamefont {{Quantum AI team and
  collaborators}}(2020{\natexlab{a}})}]{quantum_ai_team_and_collaborators_2020_4023103}%
  \BibitemOpen
  \bibfield  {author} {\bibinfo {author} {\bibnamefont {{Quantum AI team and
  collaborators}}},\ }\href {\doibase 10.5281/zenodo.4023103} {\enquote
  {\bibinfo {title} {qsim},}\ } (\bibinfo {year} {2020}{\natexlab{a}}),\
  \Eprint {http://arxiv.org/abs/https://github.com/quantumlib/qsim}
  {https://github.com/quantumlib/qsim} \BibitemShut {NoStop}%
\bibitem [{\citenamefont {Isakov}(2020)}]{isakov_blog_2020}%
  \BibitemOpen
  \bibfield  {author} {\bibinfo {author} {\bibfnamefont {Sergei}\ \bibnamefont
  {Isakov}},\ }\href
  {https://blog.google/technology/ai/qsim-explore-quantum-algorithms/}
  {\bibfield  {journal} {\bibinfo  {journal} {Researchers can use qsim to
  explore quantum algorithms, Blog Post}\ } (\bibinfo {year}
  {2020})}\BibitemShut {NoStop}%
\bibitem [{\citenamefont {{Cirq
  Developers}}(2021)}]{cirq_developers_2021_5182845}%
  \BibitemOpen
  \bibfield  {author} {\bibinfo {author} {\bibnamefont {{Cirq Developers}}},\
  }\href {\doibase 10.5281/zenodo.5182845} {\enquote {\bibinfo {title}
  {Cirq},}\ } (\bibinfo {year} {2021}),\ \bibinfo {note} {{See full list of
  authors on Github: https://github
  .com/quantumlib/Cirq/graphs/contributors}},\ \Eprint
  {http://arxiv.org/abs/https://github.com/quantumlib/Cirq}
  {https://github.com/quantumlib/Cirq} \BibitemShut {NoStop}%
\bibitem [{\citenamefont {{Quantum AI team and
  collaborators}}(2020{\natexlab{b}})}]{quantum_ai_team_and_collaborators_2020_4091470}%
  \BibitemOpen
  \bibfield  {author} {\bibinfo {author} {\bibnamefont {{Quantum AI team and
  collaborators}}},\ }\href {\doibase 10.5281/zenodo.4091470} {\enquote
  {\bibinfo {title} {Recirq},}\ } (\bibinfo {year}
  {2020}{\natexlab{b}})\BibitemShut {NoStop}%
\bibitem [{\citenamefont {McClean}\ \emph {et~al.}(2020)\citenamefont
  {McClean}, \citenamefont {Rubin}, \citenamefont {Sung}, \citenamefont
  {Kivlichan}, \citenamefont {Bonet-Monroig}, \citenamefont {Cao},
  \citenamefont {Dai}, \citenamefont {Fried}, \citenamefont {Gidney},
  \citenamefont {Gimby} \emph {et~al.}}]{mcclean2020openfermion}%
  \BibitemOpen
  \bibfield  {author} {\bibinfo {author} {\bibfnamefont {Jarrod~R}\
  \bibnamefont {McClean}}, \bibinfo {author} {\bibfnamefont {Nicholas~C}\
  \bibnamefont {Rubin}}, \bibinfo {author} {\bibfnamefont {Kevin~J}\
  \bibnamefont {Sung}}, \bibinfo {author} {\bibfnamefont {Ian~D}\ \bibnamefont
  {Kivlichan}}, \bibinfo {author} {\bibfnamefont {Xavier}\ \bibnamefont
  {Bonet-Monroig}}, \bibinfo {author} {\bibfnamefont {Yudong}\ \bibnamefont
  {Cao}}, \bibinfo {author} {\bibfnamefont {Chengyu}\ \bibnamefont {Dai}},
  \bibinfo {author} {\bibfnamefont {E~Schuyler}\ \bibnamefont {Fried}},
  \bibinfo {author} {\bibfnamefont {Craig}\ \bibnamefont {Gidney}}, \bibinfo
  {author} {\bibfnamefont {Brendan}\ \bibnamefont {Gimby}},  \emph {et~al.},\
  }\bibfield  {title} {\enquote {\bibinfo {title} {Openfermion: the electronic
  structure package for quantum computers},}\ }\href@noop {} {\bibfield
  {journal} {\bibinfo  {journal} {Quantum Science and Technology}\ }\textbf
  {\bibinfo {volume} {5}},\ \bibinfo {pages} {034014} (\bibinfo {year}
  {2020})}\BibitemShut {NoStop}%
\bibitem [{\citenamefont {Broughton}\ \emph {et~al.}(2020)\citenamefont
  {Broughton}, \citenamefont {Verdon}, \citenamefont {McCourt}, \citenamefont
  {Martinez}, \citenamefont {Yoo}, \citenamefont {Isakov}, \citenamefont
  {Massey}, \citenamefont {Halavati}, \citenamefont {Niu}, \citenamefont
  {Zlokapa} \emph {et~al.}}]{broughton2020tensorflow}%
  \BibitemOpen
  \bibfield  {author} {\bibinfo {author} {\bibfnamefont {Michael}\ \bibnamefont
  {Broughton}}, \bibinfo {author} {\bibfnamefont {Guillaume}\ \bibnamefont
  {Verdon}}, \bibinfo {author} {\bibfnamefont {Trevor}\ \bibnamefont
  {McCourt}}, \bibinfo {author} {\bibfnamefont {Antonio~J}\ \bibnamefont
  {Martinez}}, \bibinfo {author} {\bibfnamefont {Jae~Hyeon}\ \bibnamefont
  {Yoo}}, \bibinfo {author} {\bibfnamefont {Sergei~V}\ \bibnamefont {Isakov}},
  \bibinfo {author} {\bibfnamefont {Philip}\ \bibnamefont {Massey}}, \bibinfo
  {author} {\bibfnamefont {Ramin}\ \bibnamefont {Halavati}}, \bibinfo {author}
  {\bibfnamefont {Murphy~Yuezhen}\ \bibnamefont {Niu}}, \bibinfo {author}
  {\bibfnamefont {Alexander}\ \bibnamefont {Zlokapa}},  \emph {et~al.},\
  }\bibfield  {title} {\enquote {\bibinfo {title} {Tensorflow quantum: A
  software framework for quantum machine learning},}\ }\href@noop {} {\
  (\bibinfo {year} {2020})},\ \Eprint {http://arxiv.org/abs/2003.02989}
  {arXiv:2003.02989 [quant-ph]} \BibitemShut {NoStop}%
\bibitem [{\citenamefont {Suzuki}\ \emph {et~al.}(2021)\citenamefont {Suzuki},
  \citenamefont {Kawase}, \citenamefont {Masumura}, \citenamefont {Hiraga},
  \citenamefont {Nakadai}, \citenamefont {Chen}, \citenamefont {Nakanishi},
  \citenamefont {Mitarai}, \citenamefont {Imai}, \citenamefont {Tamiya} \emph
  {et~al.}}]{suzuki2021qulacs}%
  \BibitemOpen
  \bibfield  {author} {\bibinfo {author} {\bibfnamefont {Yasunari}\
  \bibnamefont {Suzuki}}, \bibinfo {author} {\bibfnamefont {Yoshiaki}\
  \bibnamefont {Kawase}}, \bibinfo {author} {\bibfnamefont {Yuya}\ \bibnamefont
  {Masumura}}, \bibinfo {author} {\bibfnamefont {Yuria}\ \bibnamefont
  {Hiraga}}, \bibinfo {author} {\bibfnamefont {Masahiro}\ \bibnamefont
  {Nakadai}}, \bibinfo {author} {\bibfnamefont {Jiabao}\ \bibnamefont {Chen}},
  \bibinfo {author} {\bibfnamefont {Ken~M}\ \bibnamefont {Nakanishi}}, \bibinfo
  {author} {\bibfnamefont {Kosuke}\ \bibnamefont {Mitarai}}, \bibinfo {author}
  {\bibfnamefont {Ryosuke}\ \bibnamefont {Imai}}, \bibinfo {author}
  {\bibfnamefont {Shiro}\ \bibnamefont {Tamiya}},  \emph {et~al.},\ }\bibfield
  {title} {\enquote {\bibinfo {title} {Qulacs: a fast and versatile quantum
  circuit simulator for research purpose},}\ }\href@noop {} {\bibfield
  {journal} {\bibinfo  {journal} {Quantum}\ }\textbf {\bibinfo {volume} {5}},\
  \bibinfo {pages} {559} (\bibinfo {year} {2021})}\BibitemShut {NoStop}%
\bibitem [{\citenamefont {Gray}(2018)}]{gray2018quimb}%
  \BibitemOpen
  \bibfield  {author} {\bibinfo {author} {\bibfnamefont {Johnnie}\ \bibnamefont
  {Gray}},\ }\bibfield  {title} {\enquote {\bibinfo {title} {quimb: A python
  package for quantum information and many-body calculations},}\ }\href@noop {}
  {\bibfield  {journal} {\bibinfo  {journal} {Journal of Open Source Software}\
  }\textbf {\bibinfo {volume} {3}},\ \bibinfo {pages} {819} (\bibinfo {year}
  {2018})}\BibitemShut {NoStop}%
\bibitem [{\citenamefont {Bergholm}\ \emph {et~al.}(2018)\citenamefont
  {Bergholm}, \citenamefont {Izaac}, \citenamefont {Schuld}, \citenamefont
  {Gogolin}, \citenamefont {Alam}, \citenamefont {Ahmed}, \citenamefont
  {Arrazola}, \citenamefont {Blank}, \citenamefont {Delgado}, \citenamefont
  {Jahangiri} \emph {et~al.}}]{bergholm2018pennylane}%
  \BibitemOpen
  \bibfield  {author} {\bibinfo {author} {\bibfnamefont {Ville}\ \bibnamefont
  {Bergholm}}, \bibinfo {author} {\bibfnamefont {Josh}\ \bibnamefont {Izaac}},
  \bibinfo {author} {\bibfnamefont {Maria}\ \bibnamefont {Schuld}}, \bibinfo
  {author} {\bibfnamefont {Christian}\ \bibnamefont {Gogolin}}, \bibinfo
  {author} {\bibfnamefont {M~Sohaib}\ \bibnamefont {Alam}}, \bibinfo {author}
  {\bibfnamefont {Shahnawaz}\ \bibnamefont {Ahmed}}, \bibinfo {author}
  {\bibfnamefont {Juan~Miguel}\ \bibnamefont {Arrazola}}, \bibinfo {author}
  {\bibfnamefont {Carsten}\ \bibnamefont {Blank}}, \bibinfo {author}
  {\bibfnamefont {Alain}\ \bibnamefont {Delgado}}, \bibinfo {author}
  {\bibfnamefont {Soran}\ \bibnamefont {Jahangiri}},  \emph {et~al.},\
  }\bibfield  {title} {\enquote {\bibinfo {title} {Pennylane: Automatic
  differentiation of hybrid quantum-classical computations},}\ }\href@noop {}
  {\  (\bibinfo {year} {2018})},\ \Eprint {http://arxiv.org/abs/1811.04968}
  {arXiv:1811.04968 [quant-ph]} \BibitemShut {NoStop}%
\bibitem [{\citenamefont {Nielsen}\ \emph {et~al.}(2020)\citenamefont
  {Nielsen}, \citenamefont {Rudinger}, \citenamefont {Proctor}, \citenamefont
  {Russo}, \citenamefont {Young},\ and\ \citenamefont
  {Blume-Kohout}}]{nielsen2020probing}%
  \BibitemOpen
  \bibfield  {author} {\bibinfo {author} {\bibfnamefont {Erik}\ \bibnamefont
  {Nielsen}}, \bibinfo {author} {\bibfnamefont {Kenneth}\ \bibnamefont
  {Rudinger}}, \bibinfo {author} {\bibfnamefont {Timothy}\ \bibnamefont
  {Proctor}}, \bibinfo {author} {\bibfnamefont {Antonio}\ \bibnamefont
  {Russo}}, \bibinfo {author} {\bibfnamefont {Kevin}\ \bibnamefont {Young}}, \
  and\ \bibinfo {author} {\bibfnamefont {Robin}\ \bibnamefont {Blume-Kohout}},\
  }\bibfield  {title} {\enquote {\bibinfo {title} {Probing quantum processor
  performance with pygsti},}\ }\href@noop {} {\bibfield  {journal} {\bibinfo
  {journal} {Quantum Science and Technology}\ }\textbf {\bibinfo {volume}
  {5}},\ \bibinfo {pages} {044002} (\bibinfo {year} {2020})}\BibitemShut
  {NoStop}%
\bibitem [{\citenamefont {Sivarajah}\ \emph {et~al.}(2020)\citenamefont
  {Sivarajah}, \citenamefont {Dilkes}, \citenamefont {Cowtan}, \citenamefont
  {Simmons}, \citenamefont {Edgington},\ and\ \citenamefont
  {Duncan}}]{sivarajah2020t}%
  \BibitemOpen
  \bibfield  {author} {\bibinfo {author} {\bibfnamefont {Seyon}\ \bibnamefont
  {Sivarajah}}, \bibinfo {author} {\bibfnamefont {Silas}\ \bibnamefont
  {Dilkes}}, \bibinfo {author} {\bibfnamefont {Alexander}\ \bibnamefont
  {Cowtan}}, \bibinfo {author} {\bibfnamefont {Will}\ \bibnamefont {Simmons}},
  \bibinfo {author} {\bibfnamefont {Alec}\ \bibnamefont {Edgington}}, \ and\
  \bibinfo {author} {\bibfnamefont {Ross}\ \bibnamefont {Duncan}},\ }\bibfield
  {title} {\enquote {\bibinfo {title} {t| ket>: a retargetable compiler for
  nisq devices},}\ }\href@noop {} {\bibfield  {journal} {\bibinfo  {journal}
  {Quantum Science and Technology}\ }\textbf {\bibinfo {volume} {6}},\ \bibinfo
  {pages} {014003} (\bibinfo {year} {2020})}\BibitemShut {NoStop}%
\bibitem [{\citenamefont {Beale}\ \emph {et~al.}(2020)\citenamefont {Beale},
  \citenamefont {Carignan-Dugas}, \citenamefont {Dahlen}, \citenamefont
  {Emerson}, \citenamefont {Hincks}, \citenamefont {Iyer}, \citenamefont
  {Jain}, \citenamefont {Hufnagel}, \citenamefont {Ospadov}, \citenamefont
  {Saunders}, \citenamefont {Stasiuk}, \citenamefont {Wallman},\ and\
  \citenamefont {Winick}}]{beale_stefanie_j_2020_3945250}%
  \BibitemOpen
  \bibfield  {author} {\bibinfo {author} {\bibfnamefont {Stefanie~J.}\
  \bibnamefont {Beale}}, \bibinfo {author} {\bibfnamefont {Arnaud}\
  \bibnamefont {Carignan-Dugas}}, \bibinfo {author} {\bibfnamefont {Dar}\
  \bibnamefont {Dahlen}}, \bibinfo {author} {\bibfnamefont {Joseph}\
  \bibnamefont {Emerson}}, \bibinfo {author} {\bibfnamefont {Ian}\ \bibnamefont
  {Hincks}}, \bibinfo {author} {\bibfnamefont {Pavithran}\ \bibnamefont
  {Iyer}}, \bibinfo {author} {\bibfnamefont {Aditya}\ \bibnamefont {Jain}},
  \bibinfo {author} {\bibfnamefont {David}\ \bibnamefont {Hufnagel}}, \bibinfo
  {author} {\bibfnamefont {Egor}\ \bibnamefont {Ospadov}}, \bibinfo {author}
  {\bibfnamefont {Jordan}\ \bibnamefont {Saunders}}, \bibinfo {author}
  {\bibfnamefont {Andrew}\ \bibnamefont {Stasiuk}}, \bibinfo {author}
  {\bibfnamefont {Joel~J.}\ \bibnamefont {Wallman}}, \ and\ \bibinfo {author}
  {\bibfnamefont {Adam}\ \bibnamefont {Winick}},\ }\href {\doibase
  10.5281/zenodo.3945250} {\enquote {\bibinfo {title} {True-q},}\ } (\bibinfo
  {year} {2020})\BibitemShut {NoStop}%
\bibitem [{\citenamefont {Smelyanskiy}\ \emph {et~al.}(2016)\citenamefont
  {Smelyanskiy}, \citenamefont {Sawaya},\ and\ \citenamefont
  {Aspuru-Guzik}}]{smelyanskiy2016}%
  \BibitemOpen
  \bibfield  {author} {\bibinfo {author} {\bibfnamefont {Mikhail}\ \bibnamefont
  {Smelyanskiy}}, \bibinfo {author} {\bibfnamefont {Nicolas P.~D.}\
  \bibnamefont {Sawaya}}, \ and\ \bibinfo {author} {\bibfnamefont {Al\'{a}n}\
  \bibnamefont {Aspuru-Guzik}},\ }\href@noop {} {\enquote {\bibinfo {title}
  {qhipster: The quantum high performance software testing environment},}\ }
  (\bibinfo {year} {2016}),\ \Eprint {http://arxiv.org/abs/1601.07195}
  {arXiv:1601.07195 [quant-ph]} \BibitemShut {NoStop}%
\bibitem [{\citenamefont {H\"{a}ner}\ and\ \citenamefont
  {Steiger}(2017)}]{haener2017}%
  \BibitemOpen
  \bibfield  {author} {\bibinfo {author} {\bibfnamefont {Thomas}\ \bibnamefont
  {H\"{a}ner}}\ and\ \bibinfo {author} {\bibfnamefont {Damian~S.}\ \bibnamefont
  {Steiger}},\ }\href@noop {} {\enquote {\bibinfo {title} {0.5 petabyte
  simulation of a 45-qubit quantum circuit},}\ } (\bibinfo {year} {2017}),\
  \Eprint {http://arxiv.org/abs/1704.01127} {arXiv:1704.01127 [quant-ph]}
  \BibitemShut {NoStop}%
\bibitem [{\citenamefont {Xing}\ and\ \citenamefont
  {Broughton}(2021)}]{xing_blog_2021}%
  \BibitemOpen
  \bibfield  {author} {\bibinfo {author} {\bibfnamefont {Cheng}\ \bibnamefont
  {Xing}}\ and\ \bibinfo {author} {\bibfnamefont {Michael}\ \bibnamefont
  {Broughton}},\ }\href
  {https://blog.google/technology/ai/qsim-explore-quantum-algorithms/}
  {\bibfield  {journal} {\bibinfo  {journal} {Training with Multiple Workers
  using TensorFlow Quantum, Blog Post}\ } (\bibinfo {year} {2021})}\BibitemShut
  {NoStop}%
\bibitem [{\citenamefont {Inc.}(2021)}]{terraform}%
  \BibitemOpen
  \bibfield  {author} {\bibinfo {author} {\bibfnamefont {Hashicorp}\
  \bibnamefont {Inc.}},\ }\href {https://www.terraform.io/} {\enquote {\bibinfo
  {title} {Terraform by hashicorp},}\ } (\bibinfo {year} {2021}),\ \bibinfo
  {note} {{https://www.terraform.io/}}\BibitemShut {NoStop}%
\bibitem [{\citenamefont {Thain}\ \emph {et~al.}(2005)\citenamefont {Thain},
  \citenamefont {Tannenbaum},\ and\ \citenamefont {Livny}}]{htcondor}%
  \BibitemOpen
  \bibfield  {author} {\bibinfo {author} {\bibfnamefont {Douglas}\ \bibnamefont
  {Thain}}, \bibinfo {author} {\bibfnamefont {Todd}\ \bibnamefont
  {Tannenbaum}}, \ and\ \bibinfo {author} {\bibfnamefont {Miron}\ \bibnamefont
  {Livny}},\ }\bibfield  {title} {\enquote {\bibinfo {title} {Distributed
  computing in practice: the condor experience.}}\ }\href@noop {} {\bibfield
  {journal} {\bibinfo  {journal} {Concurrency - Practice and Experience}\
  }\textbf {\bibinfo {volume} {17}},\ \bibinfo {pages} {323--356} (\bibinfo
  {year} {2005})}\BibitemShut {NoStop}%
\bibitem [{\citenamefont {Gustafson}\ \emph {et~al.}(2021)\citenamefont
  {Gustafson}, \citenamefont {Holzman}, \citenamefont {Kowalkowski},
  \citenamefont {Lamm}, \citenamefont {Li}, \citenamefont {Perdue},
  \citenamefont {Boixo}, \citenamefont {Isakov}, \citenamefont {Martin},
  \citenamefont {Thomson}, \citenamefont {Heidweiller}, \citenamefont {Beall},
  \citenamefont {Ganahl}, \citenamefont {Vidal},\ and\ \citenamefont
  {Peters}}]{gustafson2021large}%
  \BibitemOpen
  \bibfield  {author} {\bibinfo {author} {\bibfnamefont {Erik}\ \bibnamefont
  {Gustafson}}, \bibinfo {author} {\bibfnamefont {Burt}\ \bibnamefont
  {Holzman}}, \bibinfo {author} {\bibfnamefont {James}\ \bibnamefont
  {Kowalkowski}}, \bibinfo {author} {\bibfnamefont {Henry}\ \bibnamefont
  {Lamm}}, \bibinfo {author} {\bibfnamefont {Andy C.~Y.}\ \bibnamefont {Li}},
  \bibinfo {author} {\bibfnamefont {Gabriel}\ \bibnamefont {Perdue}}, \bibinfo
  {author} {\bibfnamefont {Sergio}\ \bibnamefont {Boixo}}, \bibinfo {author}
  {\bibfnamefont {Sergei~V.}\ \bibnamefont {Isakov}}, \bibinfo {author}
  {\bibfnamefont {Orion}\ \bibnamefont {Martin}}, \bibinfo {author}
  {\bibfnamefont {Ross}\ \bibnamefont {Thomson}}, \bibinfo {author}
  {\bibfnamefont {Catherine~Vollgraff}\ \bibnamefont {Heidweiller}}, \bibinfo
  {author} {\bibfnamefont {Jackson}\ \bibnamefont {Beall}}, \bibinfo {author}
  {\bibfnamefont {Martin}\ \bibnamefont {Ganahl}}, \bibinfo {author}
  {\bibfnamefont {Guifre}\ \bibnamefont {Vidal}}, \ and\ \bibinfo {author}
  {\bibfnamefont {Evan}\ \bibnamefont {Peters}},\ }\href@noop {} {\enquote
  {\bibinfo {title} {Large scale multi-node simulations of $\mathbb{Z}_2$ gauge
  theory quantum circuits using google cloud platform},}\ } (\bibinfo {year}
  {2021}),\ \Eprint {http://arxiv.org/abs/2110.07482} {arXiv:2110.07482
  [quant-ph]} \BibitemShut {NoStop}%
\bibitem [{\citenamefont {Feng}\ \emph {et~al.}(2016)\citenamefont {Feng},
  \citenamefont {Wallman}, \citenamefont {Buonacorsi}, \citenamefont {Cho},
  \citenamefont {Park}, \citenamefont {Xin}, \citenamefont {Lu}, \citenamefont
  {Baugh},\ and\ \citenamefont {Laflamme}}]{feng2016estimating}%
  \BibitemOpen
  \bibfield  {author} {\bibinfo {author} {\bibfnamefont {Guanru}\ \bibnamefont
  {Feng}}, \bibinfo {author} {\bibfnamefont {Joel~J}\ \bibnamefont {Wallman}},
  \bibinfo {author} {\bibfnamefont {Brandon}\ \bibnamefont {Buonacorsi}},
  \bibinfo {author} {\bibfnamefont {Franklin~H}\ \bibnamefont {Cho}}, \bibinfo
  {author} {\bibfnamefont {Daniel~K}\ \bibnamefont {Park}}, \bibinfo {author}
  {\bibfnamefont {Tao}\ \bibnamefont {Xin}}, \bibinfo {author} {\bibfnamefont
  {Dawei}\ \bibnamefont {Lu}}, \bibinfo {author} {\bibfnamefont {Jonathan}\
  \bibnamefont {Baugh}}, \ and\ \bibinfo {author} {\bibfnamefont {Raymond}\
  \bibnamefont {Laflamme}},\ }\bibfield  {title} {\enquote {\bibinfo {title}
  {Estimating the coherence of noise in quantum control of a solid-state
  qubit},}\ }\href@noop {} {\bibfield  {journal} {\bibinfo  {journal} {Physical
  Review Letters}\ }\textbf {\bibinfo {volume} {117}},\ \bibinfo {pages}
  {260501} (\bibinfo {year} {2016})}\BibitemShut {NoStop}%
\bibitem [{\citenamefont {Magesan}\ \emph {et~al.}(2011)\citenamefont
  {Magesan}, \citenamefont {Gambetta},\ and\ \citenamefont
  {Emerson}}]{magesan2011scalable}%
  \BibitemOpen
  \bibfield  {author} {\bibinfo {author} {\bibfnamefont {Easwar}\ \bibnamefont
  {Magesan}}, \bibinfo {author} {\bibfnamefont {Jay~M}\ \bibnamefont
  {Gambetta}}, \ and\ \bibinfo {author} {\bibfnamefont {Joseph}\ \bibnamefont
  {Emerson}},\ }\bibfield  {title} {\enquote {\bibinfo {title} {Scalable and
  robust randomized benchmarking of quantum processes},}\ }\href@noop {}
  {\bibfield  {journal} {\bibinfo  {journal} {Physical review letters}\
  }\textbf {\bibinfo {volume} {106}},\ \bibinfo {pages} {180504} (\bibinfo
  {year} {2011})}\BibitemShut {NoStop}%
\bibitem [{\citenamefont {Magesan}\ \emph {et~al.}(2012)\citenamefont
  {Magesan}, \citenamefont {Gambetta},\ and\ \citenamefont
  {Emerson}}]{magesan2012characterizing}%
  \BibitemOpen
  \bibfield  {author} {\bibinfo {author} {\bibfnamefont {Easwar}\ \bibnamefont
  {Magesan}}, \bibinfo {author} {\bibfnamefont {Jay~M}\ \bibnamefont
  {Gambetta}}, \ and\ \bibinfo {author} {\bibfnamefont {Joseph}\ \bibnamefont
  {Emerson}},\ }\bibfield  {title} {\enquote {\bibinfo {title} {Characterizing
  quantum gates via randomized benchmarking},}\ }\href@noop {} {\bibfield
  {journal} {\bibinfo  {journal} {Physical Review A}\ }\textbf {\bibinfo
  {volume} {85}},\ \bibinfo {pages} {042311} (\bibinfo {year}
  {2012})}\BibitemShut {NoStop}%
\bibitem [{\citenamefont {Boixo}\ \emph {et~al.}(2018)\citenamefont {Boixo},
  \citenamefont {Isakov}, \citenamefont {Smelyanskiy}, \citenamefont {Babbush},
  \citenamefont {Ding}, \citenamefont {Jiang}, \citenamefont {Bremner},
  \citenamefont {Martinis},\ and\ \citenamefont
  {Neven}}]{boixo2018characterizing}%
  \BibitemOpen
  \bibfield  {author} {\bibinfo {author} {\bibfnamefont {Sergio}\ \bibnamefont
  {Boixo}}, \bibinfo {author} {\bibfnamefont {Sergei~V}\ \bibnamefont
  {Isakov}}, \bibinfo {author} {\bibfnamefont {Vadim~N}\ \bibnamefont
  {Smelyanskiy}}, \bibinfo {author} {\bibfnamefont {Ryan}\ \bibnamefont
  {Babbush}}, \bibinfo {author} {\bibfnamefont {Nan}\ \bibnamefont {Ding}},
  \bibinfo {author} {\bibfnamefont {Zhang}\ \bibnamefont {Jiang}}, \bibinfo
  {author} {\bibfnamefont {Michael~J}\ \bibnamefont {Bremner}}, \bibinfo
  {author} {\bibfnamefont {John~M}\ \bibnamefont {Martinis}}, \ and\ \bibinfo
  {author} {\bibfnamefont {Hartmut}\ \bibnamefont {Neven}},\ }\bibfield
  {title} {\enquote {\bibinfo {title} {Characterizing quantum supremacy in
  near-term devices},}\ }\href@noop {} {\bibfield  {journal} {\bibinfo
  {journal} {Nature Physics}\ }\textbf {\bibinfo {volume} {14}},\ \bibinfo
  {pages} {595--600} (\bibinfo {year} {2018})}\BibitemShut {NoStop}%
\bibitem [{\citenamefont {Neill}\ \emph {et~al.}(2018)\citenamefont {Neill},
  \citenamefont {Roushan}, \citenamefont {Kechedzhi}, \citenamefont {Boixo},
  \citenamefont {Isakov}, \citenamefont {Smelyanskiy}, \citenamefont {Megrant},
  \citenamefont {Chiaro}, \citenamefont {Dunsworth}, \citenamefont {Arya} \emph
  {et~al.}}]{neill2018blueprint}%
  \BibitemOpen
  \bibfield  {author} {\bibinfo {author} {\bibfnamefont {Charles}\ \bibnamefont
  {Neill}}, \bibinfo {author} {\bibfnamefont {Pedran}\ \bibnamefont {Roushan}},
  \bibinfo {author} {\bibfnamefont {K}~\bibnamefont {Kechedzhi}}, \bibinfo
  {author} {\bibfnamefont {Sergio}\ \bibnamefont {Boixo}}, \bibinfo {author}
  {\bibfnamefont {Sergei~V}\ \bibnamefont {Isakov}}, \bibinfo {author}
  {\bibfnamefont {V}~\bibnamefont {Smelyanskiy}}, \bibinfo {author}
  {\bibfnamefont {A}~\bibnamefont {Megrant}}, \bibinfo {author} {\bibfnamefont
  {B}~\bibnamefont {Chiaro}}, \bibinfo {author} {\bibfnamefont {A}~\bibnamefont
  {Dunsworth}}, \bibinfo {author} {\bibfnamefont {K}~\bibnamefont {Arya}},
  \emph {et~al.},\ }\bibfield  {title} {\enquote {\bibinfo {title} {A blueprint
  for demonstrating quantum supremacy with superconducting qubits},}\
  }\href@noop {} {\bibfield  {journal} {\bibinfo  {journal} {Science}\ }\textbf
  {\bibinfo {volume} {360}},\ \bibinfo {pages} {195--199} (\bibinfo {year}
  {2018})}\BibitemShut {NoStop}%
\bibitem [{\citenamefont {Arute}\ \emph
  {et~al.}(2019{\natexlab{a}})\citenamefont {Arute}, \citenamefont {Arya},
  \citenamefont {Babbush}, \citenamefont {Bacon}, \citenamefont {Bardin},
  \citenamefont {Barends}, \citenamefont {Biswas}, \citenamefont {Boixo},
  \citenamefont {Brandao}, \citenamefont {Buell} \emph
  {et~al.}}]{arute2019quantum}%
  \BibitemOpen
  \bibfield  {author} {\bibinfo {author} {\bibfnamefont {Frank}\ \bibnamefont
  {Arute}}, \bibinfo {author} {\bibfnamefont {Kunal}\ \bibnamefont {Arya}},
  \bibinfo {author} {\bibfnamefont {Ryan}\ \bibnamefont {Babbush}}, \bibinfo
  {author} {\bibfnamefont {Dave}\ \bibnamefont {Bacon}}, \bibinfo {author}
  {\bibfnamefont {Joseph~C}\ \bibnamefont {Bardin}}, \bibinfo {author}
  {\bibfnamefont {Rami}\ \bibnamefont {Barends}}, \bibinfo {author}
  {\bibfnamefont {Rupak}\ \bibnamefont {Biswas}}, \bibinfo {author}
  {\bibfnamefont {Sergio}\ \bibnamefont {Boixo}}, \bibinfo {author}
  {\bibfnamefont {Fernando~GSL}\ \bibnamefont {Brandao}}, \bibinfo {author}
  {\bibfnamefont {David~A}\ \bibnamefont {Buell}},  \emph {et~al.},\ }\bibfield
   {title} {\enquote {\bibinfo {title} {Quantum supremacy using a programmable
  superconducting processor},}\ }\href@noop {} {\bibfield  {journal} {\bibinfo
  {journal} {Nature}\ }\textbf {\bibinfo {volume} {574}},\ \bibinfo {pages}
  {505--510} (\bibinfo {year} {2019}{\natexlab{a}})}\BibitemShut {NoStop}%
\bibitem [{\citenamefont {Nielsen}\ and\ \citenamefont
  {Chuang}(2002)}]{nielsen2002quantum}%
  \BibitemOpen
  \bibfield  {author} {\bibinfo {author} {\bibfnamefont {Michael~A}\
  \bibnamefont {Nielsen}}\ and\ \bibinfo {author} {\bibfnamefont {Isaac}\
  \bibnamefont {Chuang}},\ }\href@noop {} {\enquote {\bibinfo {title} {Quantum
  computation and quantum information},}\ } (\bibinfo {year}
  {2002})\BibitemShut {NoStop}%
\bibitem [{\citenamefont {Foxen}\ \emph {et~al.}(2020)\citenamefont {Foxen},
  \citenamefont {Neill}, \citenamefont {Dunsworth}, \citenamefont {Roushan},
  \citenamefont {Chiaro}, \citenamefont {Megrant}, \citenamefont {Kelly},
  \citenamefont {Chen}, \citenamefont {Satzinger}, \citenamefont {Barends},\
  and\ \citenamefont {et~al.}}]{2020}%
  \BibitemOpen
  \bibfield  {author} {\bibinfo {author} {\bibfnamefont {B.}~\bibnamefont
  {Foxen}}, \bibinfo {author} {\bibfnamefont {C.}~\bibnamefont {Neill}},
  \bibinfo {author} {\bibfnamefont {A.}~\bibnamefont {Dunsworth}}, \bibinfo
  {author} {\bibfnamefont {P.}~\bibnamefont {Roushan}}, \bibinfo {author}
  {\bibfnamefont {B.}~\bibnamefont {Chiaro}}, \bibinfo {author} {\bibfnamefont
  {A.}~\bibnamefont {Megrant}}, \bibinfo {author} {\bibfnamefont
  {J.}~\bibnamefont {Kelly}}, \bibinfo {author} {\bibfnamefont {Zijun}\
  \bibnamefont {Chen}}, \bibinfo {author} {\bibfnamefont {K.}~\bibnamefont
  {Satzinger}}, \bibinfo {author} {\bibfnamefont {R.}~\bibnamefont {Barends}},
  \ and\ \bibinfo {author} {\bibnamefont {et~al.}},\ }\bibfield  {title}
  {\enquote {\bibinfo {title} {Demonstrating a continuous set of two-qubit
  gates for near-term quantum algorithms},}\ }\href
  {http://dx.doi.org/10.1103/PhysRevLett.125.120504} {\bibfield  {journal}
  {\bibinfo  {journal} {Physical Review Letters}\ }\textbf {\bibinfo {volume}
  {125}} (\bibinfo {year} {2020})}\BibitemShut {NoStop}%
\bibitem [{\citenamefont {McEwen}\ \emph {et~al.}(2021)\citenamefont {McEwen},
  \citenamefont {Kafri}, \citenamefont {Chen}, \citenamefont {Atalaya},
  \citenamefont {Satzinger}, \citenamefont {Quintana}, \citenamefont {Klimov},
  \citenamefont {Sank}, \citenamefont {Gidney}, \citenamefont {Fowler},
  \citenamefont {Arute}, \citenamefont {Arya}, \citenamefont {Buckley},
  \citenamefont {Burkett}, \citenamefont {Bushnell}, \citenamefont {Chiaro},
  \citenamefont {Collins}, \citenamefont {Demura}, \citenamefont {Dunsworth},
  \citenamefont {Erickson}, \citenamefont {Foxen}, \citenamefont {Giustina},
  \citenamefont {Huang}, \citenamefont {Hong}, \citenamefont {Jeffrey},
  \citenamefont {Kim}, \citenamefont {Kechedzhi}, \citenamefont {Kostritsa},
  \citenamefont {Laptev}, \citenamefont {Megrant}, \citenamefont {Mi},
  \citenamefont {Mutus}, \citenamefont {Naaman}, \citenamefont {Neeley},
  \citenamefont {Neill}, \citenamefont {Niu}, \citenamefont {Paler},
  \citenamefont {Redd}, \citenamefont {Roushan}, \citenamefont {White},
  \citenamefont {Yao}, \citenamefont {Yeh}, \citenamefont {Zalcman},
  \citenamefont {Chen}, \citenamefont {Smelyanskiy}, \citenamefont {Martinis},
  \citenamefont {Neven}, \citenamefont {Kelly}, \citenamefont {Korotkov},
  \citenamefont {Petukhov},\ and\ \citenamefont {Barends}}]{2021}%
  \BibitemOpen
  \bibfield  {author} {\bibinfo {author} {\bibfnamefont {M.}~\bibnamefont
  {McEwen}}, \bibinfo {author} {\bibfnamefont {D.}~\bibnamefont {Kafri}},
  \bibinfo {author} {\bibfnamefont {Z.}~\bibnamefont {Chen}}, \bibinfo {author}
  {\bibfnamefont {J.}~\bibnamefont {Atalaya}}, \bibinfo {author} {\bibfnamefont
  {K.~J.}\ \bibnamefont {Satzinger}}, \bibinfo {author} {\bibfnamefont
  {C.}~\bibnamefont {Quintana}}, \bibinfo {author} {\bibfnamefont {P.~V.}\
  \bibnamefont {Klimov}}, \bibinfo {author} {\bibfnamefont {D.}~\bibnamefont
  {Sank}}, \bibinfo {author} {\bibfnamefont {C.}~\bibnamefont {Gidney}},
  \bibinfo {author} {\bibfnamefont {A.~G.}\ \bibnamefont {Fowler}}, \bibinfo
  {author} {\bibfnamefont {F.}~\bibnamefont {Arute}}, \bibinfo {author}
  {\bibfnamefont {K.}~\bibnamefont {Arya}}, \bibinfo {author} {\bibfnamefont
  {B.}~\bibnamefont {Buckley}}, \bibinfo {author} {\bibfnamefont
  {B.}~\bibnamefont {Burkett}}, \bibinfo {author} {\bibfnamefont
  {N.}~\bibnamefont {Bushnell}}, \bibinfo {author} {\bibfnamefont
  {B.}~\bibnamefont {Chiaro}}, \bibinfo {author} {\bibfnamefont
  {R.}~\bibnamefont {Collins}}, \bibinfo {author} {\bibfnamefont
  {S.}~\bibnamefont {Demura}}, \bibinfo {author} {\bibfnamefont
  {A.}~\bibnamefont {Dunsworth}}, \bibinfo {author} {\bibfnamefont
  {C.}~\bibnamefont {Erickson}}, \bibinfo {author} {\bibfnamefont
  {B.}~\bibnamefont {Foxen}}, \bibinfo {author} {\bibfnamefont
  {M.}~\bibnamefont {Giustina}}, \bibinfo {author} {\bibfnamefont
  {T.}~\bibnamefont {Huang}}, \bibinfo {author} {\bibfnamefont
  {S.}~\bibnamefont {Hong}}, \bibinfo {author} {\bibfnamefont {E.}~\bibnamefont
  {Jeffrey}}, \bibinfo {author} {\bibfnamefont {S.}~\bibnamefont {Kim}},
  \bibinfo {author} {\bibfnamefont {K.}~\bibnamefont {Kechedzhi}}, \bibinfo
  {author} {\bibfnamefont {F.}~\bibnamefont {Kostritsa}}, \bibinfo {author}
  {\bibfnamefont {P.}~\bibnamefont {Laptev}}, \bibinfo {author} {\bibfnamefont
  {A.}~\bibnamefont {Megrant}}, \bibinfo {author} {\bibfnamefont
  {X.}~\bibnamefont {Mi}}, \bibinfo {author} {\bibfnamefont {J.}~\bibnamefont
  {Mutus}}, \bibinfo {author} {\bibfnamefont {O.}~\bibnamefont {Naaman}},
  \bibinfo {author} {\bibfnamefont {M.}~\bibnamefont {Neeley}}, \bibinfo
  {author} {\bibfnamefont {C.}~\bibnamefont {Neill}}, \bibinfo {author}
  {\bibfnamefont {M.}~\bibnamefont {Niu}}, \bibinfo {author} {\bibfnamefont
  {A.}~\bibnamefont {Paler}}, \bibinfo {author} {\bibfnamefont
  {N.}~\bibnamefont {Redd}}, \bibinfo {author} {\bibfnamefont {P.}~\bibnamefont
  {Roushan}}, \bibinfo {author} {\bibfnamefont {T.~C.}\ \bibnamefont {White}},
  \bibinfo {author} {\bibfnamefont {J.}~\bibnamefont {Yao}}, \bibinfo {author}
  {\bibfnamefont {P.}~\bibnamefont {Yeh}}, \bibinfo {author} {\bibfnamefont
  {A.}~\bibnamefont {Zalcman}}, \bibinfo {author} {\bibfnamefont
  {Yu}~\bibnamefont {Chen}}, \bibinfo {author} {\bibfnamefont {V.~N.}\
  \bibnamefont {Smelyanskiy}}, \bibinfo {author} {\bibfnamefont {John~M.}\
  \bibnamefont {Martinis}}, \bibinfo {author} {\bibfnamefont {H.}~\bibnamefont
  {Neven}}, \bibinfo {author} {\bibfnamefont {J.}~\bibnamefont {Kelly}},
  \bibinfo {author} {\bibfnamefont {A.~N.}\ \bibnamefont {Korotkov}}, \bibinfo
  {author} {\bibfnamefont {A.~G.}\ \bibnamefont {Petukhov}}, \ and\ \bibinfo
  {author} {\bibfnamefont {R.}~\bibnamefont {Barends}},\ }\bibfield  {title}
  {\enquote {\bibinfo {title} {Removing leakage-induced correlated errors in
  superconducting quantum error correction},}\ }\href {\doibase
  10.1038/s41467-021-21982-y} {\ \textbf {\bibinfo {volume} {12}} (\bibinfo
  {year} {2021}),\ 10.1038/s41467-021-21982-y}\BibitemShut {NoStop}%
\bibitem [{\citenamefont {Arute}\ \emph
  {et~al.}(2020{\natexlab{a}})\citenamefont {Arute}, \citenamefont {Arya},
  \citenamefont {Babbush}, \citenamefont {Bacon}, \citenamefont {Bardin},
  \citenamefont {Barends}, \citenamefont {Bengtsson}, \citenamefont {Boixo},
  \citenamefont {Broughton}, \citenamefont {Buckley}, \citenamefont {Buell},
  \citenamefont {Burkett}, \citenamefont {Bushnell}, \citenamefont {Chen},
  \citenamefont {Chen}, \citenamefont {Chen}, \citenamefont {Chiaro},
  \citenamefont {Collins}, \citenamefont {Cotton}, \citenamefont {Courtney},
  \citenamefont {Demura}, \citenamefont {Derk}, \citenamefont {Dunsworth},
  \citenamefont {Eppens}, \citenamefont {Eckl}, \citenamefont {Erickson},
  \citenamefont {Farhi}, \citenamefont {Fowler}, \citenamefont {Foxen},
  \citenamefont {Gidney}, \citenamefont {Giustina}, \citenamefont {Graff},
  \citenamefont {Gross}, \citenamefont {Habegger}, \citenamefont {Harrigan},
  \citenamefont {Ho}, \citenamefont {Hong}, \citenamefont {Huang},
  \citenamefont {Huggins}, \citenamefont {Ioffe}, \citenamefont {Isakov},
  \citenamefont {Jeffrey}, \citenamefont {Jiang}, \citenamefont {Jones},
  \citenamefont {Kafri}, \citenamefont {Kechedzhi}, \citenamefont {Kelly},
  \citenamefont {Kim}, \citenamefont {Klimov}, \citenamefont {Korotkov},
  \citenamefont {Kostritsa}, \citenamefont {Landhuis}, \citenamefont {Laptev},
  \citenamefont {Lindmark}, \citenamefont {Lucero}, \citenamefont {Marthaler},
  \citenamefont {Martin}, \citenamefont {Martinis}, \citenamefont {Marusczyk},
  \citenamefont {McArdle}, \citenamefont {McClean}, \citenamefont {McCourt},
  \citenamefont {McEwen}, \citenamefont {Megrant}, \citenamefont
  {Mejuto-Zaera}, \citenamefont {Mi}, \citenamefont {Mohseni}, \citenamefont
  {Mruczkiewicz}, \citenamefont {Mutus}, \citenamefont {Naaman}, \citenamefont
  {Neeley}, \citenamefont {Neill}, \citenamefont {Neven}, \citenamefont
  {Newman}, \citenamefont {Niu}, \citenamefont {O'Brien}, \citenamefont
  {Ostby}, \citenamefont {Pató}, \citenamefont {Petukhov}, \citenamefont
  {Putterman}, \citenamefont {Quintana}, \citenamefont {Reiner}, \citenamefont
  {Roushan}, \citenamefont {Rubin}, \citenamefont {Sank}, \citenamefont
  {Satzinger}, \citenamefont {Smelyanskiy}, \citenamefont {Strain},
  \citenamefont {Sung}, \citenamefont {Schmitteckert}, \citenamefont {Szalay},
  \citenamefont {Tubman}, \citenamefont {Vainsencher}, \citenamefont {White},
  \citenamefont {Vogt}, \citenamefont {Yao}, \citenamefont {Yeh}, \citenamefont
  {Zalcman},\ and\ \citenamefont {Zanker}}]{arute2020observation}%
  \BibitemOpen
  \bibfield  {author} {\bibinfo {author} {\bibfnamefont {Frank}\ \bibnamefont
  {Arute}}, \bibinfo {author} {\bibfnamefont {Kunal}\ \bibnamefont {Arya}},
  \bibinfo {author} {\bibfnamefont {Ryan}\ \bibnamefont {Babbush}}, \bibinfo
  {author} {\bibfnamefont {Dave}\ \bibnamefont {Bacon}}, \bibinfo {author}
  {\bibfnamefont {Joseph~C.}\ \bibnamefont {Bardin}}, \bibinfo {author}
  {\bibfnamefont {Rami}\ \bibnamefont {Barends}}, \bibinfo {author}
  {\bibfnamefont {Andreas}\ \bibnamefont {Bengtsson}}, \bibinfo {author}
  {\bibfnamefont {Sergio}\ \bibnamefont {Boixo}}, \bibinfo {author}
  {\bibfnamefont {Michael}\ \bibnamefont {Broughton}}, \bibinfo {author}
  {\bibfnamefont {Bob~B.}\ \bibnamefont {Buckley}}, \bibinfo {author}
  {\bibfnamefont {David~A.}\ \bibnamefont {Buell}}, \bibinfo {author}
  {\bibfnamefont {Brian}\ \bibnamefont {Burkett}}, \bibinfo {author}
  {\bibfnamefont {Nicholas}\ \bibnamefont {Bushnell}}, \bibinfo {author}
  {\bibfnamefont {Yu}~\bibnamefont {Chen}}, \bibinfo {author} {\bibfnamefont
  {Zijun}\ \bibnamefont {Chen}}, \bibinfo {author} {\bibfnamefont {Yu-An}\
  \bibnamefont {Chen}}, \bibinfo {author} {\bibfnamefont {Ben}\ \bibnamefont
  {Chiaro}}, \bibinfo {author} {\bibfnamefont {Roberto}\ \bibnamefont
  {Collins}}, \bibinfo {author} {\bibfnamefont {Stephen~J.}\ \bibnamefont
  {Cotton}}, \bibinfo {author} {\bibfnamefont {William}\ \bibnamefont
  {Courtney}}, \bibinfo {author} {\bibfnamefont {Sean}\ \bibnamefont {Demura}},
  \bibinfo {author} {\bibfnamefont {Alan}\ \bibnamefont {Derk}}, \bibinfo
  {author} {\bibfnamefont {Andrew}\ \bibnamefont {Dunsworth}}, \bibinfo
  {author} {\bibfnamefont {Daniel}\ \bibnamefont {Eppens}}, \bibinfo {author}
  {\bibfnamefont {Thomas}\ \bibnamefont {Eckl}}, \bibinfo {author}
  {\bibfnamefont {Catherine}\ \bibnamefont {Erickson}}, \bibinfo {author}
  {\bibfnamefont {Edward}\ \bibnamefont {Farhi}}, \bibinfo {author}
  {\bibfnamefont {Austin}\ \bibnamefont {Fowler}}, \bibinfo {author}
  {\bibfnamefont {Brooks}\ \bibnamefont {Foxen}}, \bibinfo {author}
  {\bibfnamefont {Craig}\ \bibnamefont {Gidney}}, \bibinfo {author}
  {\bibfnamefont {Marissa}\ \bibnamefont {Giustina}}, \bibinfo {author}
  {\bibfnamefont {Rob}\ \bibnamefont {Graff}}, \bibinfo {author} {\bibfnamefont
  {Jonathan~A.}\ \bibnamefont {Gross}}, \bibinfo {author} {\bibfnamefont
  {Steve}\ \bibnamefont {Habegger}}, \bibinfo {author} {\bibfnamefont
  {Matthew~P.}\ \bibnamefont {Harrigan}}, \bibinfo {author} {\bibfnamefont
  {Alan}\ \bibnamefont {Ho}}, \bibinfo {author} {\bibfnamefont {Sabrina}\
  \bibnamefont {Hong}}, \bibinfo {author} {\bibfnamefont {Trent}\ \bibnamefont
  {Huang}}, \bibinfo {author} {\bibfnamefont {William}\ \bibnamefont
  {Huggins}}, \bibinfo {author} {\bibfnamefont {Lev~B.}\ \bibnamefont {Ioffe}},
  \bibinfo {author} {\bibfnamefont {Sergei~V.}\ \bibnamefont {Isakov}},
  \bibinfo {author} {\bibfnamefont {Evan}\ \bibnamefont {Jeffrey}}, \bibinfo
  {author} {\bibfnamefont {Zhang}\ \bibnamefont {Jiang}}, \bibinfo {author}
  {\bibfnamefont {Cody}\ \bibnamefont {Jones}}, \bibinfo {author}
  {\bibfnamefont {Dvir}\ \bibnamefont {Kafri}}, \bibinfo {author}
  {\bibfnamefont {Kostyantyn}\ \bibnamefont {Kechedzhi}}, \bibinfo {author}
  {\bibfnamefont {Julian}\ \bibnamefont {Kelly}}, \bibinfo {author}
  {\bibfnamefont {Seon}\ \bibnamefont {Kim}}, \bibinfo {author} {\bibfnamefont
  {Paul~V.}\ \bibnamefont {Klimov}}, \bibinfo {author} {\bibfnamefont
  {Alexander~N.}\ \bibnamefont {Korotkov}}, \bibinfo {author} {\bibfnamefont
  {Fedor}\ \bibnamefont {Kostritsa}}, \bibinfo {author} {\bibfnamefont {David}\
  \bibnamefont {Landhuis}}, \bibinfo {author} {\bibfnamefont {Pavel}\
  \bibnamefont {Laptev}}, \bibinfo {author} {\bibfnamefont {Mike}\ \bibnamefont
  {Lindmark}}, \bibinfo {author} {\bibfnamefont {Erik}\ \bibnamefont {Lucero}},
  \bibinfo {author} {\bibfnamefont {Michael}\ \bibnamefont {Marthaler}},
  \bibinfo {author} {\bibfnamefont {Orion}\ \bibnamefont {Martin}}, \bibinfo
  {author} {\bibfnamefont {John~M.}\ \bibnamefont {Martinis}}, \bibinfo
  {author} {\bibfnamefont {Anika}\ \bibnamefont {Marusczyk}}, \bibinfo {author}
  {\bibfnamefont {Sam}\ \bibnamefont {McArdle}}, \bibinfo {author}
  {\bibfnamefont {Jarrod~R.}\ \bibnamefont {McClean}}, \bibinfo {author}
  {\bibfnamefont {Trevor}\ \bibnamefont {McCourt}}, \bibinfo {author}
  {\bibfnamefont {Matt}\ \bibnamefont {McEwen}}, \bibinfo {author}
  {\bibfnamefont {Anthony}\ \bibnamefont {Megrant}}, \bibinfo {author}
  {\bibfnamefont {Carlos}\ \bibnamefont {Mejuto-Zaera}}, \bibinfo {author}
  {\bibfnamefont {Xiao}\ \bibnamefont {Mi}}, \bibinfo {author} {\bibfnamefont
  {Masoud}\ \bibnamefont {Mohseni}}, \bibinfo {author} {\bibfnamefont
  {Wojciech}\ \bibnamefont {Mruczkiewicz}}, \bibinfo {author} {\bibfnamefont
  {Josh}\ \bibnamefont {Mutus}}, \bibinfo {author} {\bibfnamefont {Ofer}\
  \bibnamefont {Naaman}}, \bibinfo {author} {\bibfnamefont {Matthew}\
  \bibnamefont {Neeley}}, \bibinfo {author} {\bibfnamefont {Charles}\
  \bibnamefont {Neill}}, \bibinfo {author} {\bibfnamefont {Hartmut}\
  \bibnamefont {Neven}}, \bibinfo {author} {\bibfnamefont {Michael}\
  \bibnamefont {Newman}}, \bibinfo {author} {\bibfnamefont {Murphy~Yuezhen}\
  \bibnamefont {Niu}}, \bibinfo {author} {\bibfnamefont {Thomas~E.}\
  \bibnamefont {O'Brien}}, \bibinfo {author} {\bibfnamefont {Eric}\
  \bibnamefont {Ostby}}, \bibinfo {author} {\bibfnamefont {Bálint}\
  \bibnamefont {Pató}}, \bibinfo {author} {\bibfnamefont {Andre}\ \bibnamefont
  {Petukhov}}, \bibinfo {author} {\bibfnamefont {Harald}\ \bibnamefont
  {Putterman}}, \bibinfo {author} {\bibfnamefont {Chris}\ \bibnamefont
  {Quintana}}, \bibinfo {author} {\bibfnamefont {Jan-Michael}\ \bibnamefont
  {Reiner}}, \bibinfo {author} {\bibfnamefont {Pedram}\ \bibnamefont
  {Roushan}}, \bibinfo {author} {\bibfnamefont {Nicholas~C.}\ \bibnamefont
  {Rubin}}, \bibinfo {author} {\bibfnamefont {Daniel}\ \bibnamefont {Sank}},
  \bibinfo {author} {\bibfnamefont {Kevin~J.}\ \bibnamefont {Satzinger}},
  \bibinfo {author} {\bibfnamefont {Vadim}\ \bibnamefont {Smelyanskiy}},
  \bibinfo {author} {\bibfnamefont {Doug}\ \bibnamefont {Strain}}, \bibinfo
  {author} {\bibfnamefont {Kevin~J.}\ \bibnamefont {Sung}}, \bibinfo {author}
  {\bibfnamefont {Peter}\ \bibnamefont {Schmitteckert}}, \bibinfo {author}
  {\bibfnamefont {Marco}\ \bibnamefont {Szalay}}, \bibinfo {author}
  {\bibfnamefont {Norm~M.}\ \bibnamefont {Tubman}}, \bibinfo {author}
  {\bibfnamefont {Amit}\ \bibnamefont {Vainsencher}}, \bibinfo {author}
  {\bibfnamefont {Theodore}\ \bibnamefont {White}}, \bibinfo {author}
  {\bibfnamefont {Nicolas}\ \bibnamefont {Vogt}}, \bibinfo {author}
  {\bibfnamefont {Z.~Jamie}\ \bibnamefont {Yao}}, \bibinfo {author}
  {\bibfnamefont {Ping}\ \bibnamefont {Yeh}}, \bibinfo {author} {\bibfnamefont
  {Adam}\ \bibnamefont {Zalcman}}, \ and\ \bibinfo {author} {\bibfnamefont
  {Sebastian}\ \bibnamefont {Zanker}},\ }\href@noop {} {\enquote {\bibinfo
  {title} {Observation of separated dynamics of charge and spin in the
  fermi-hubbard model},}\ } (\bibinfo {year} {2020}{\natexlab{a}}),\ \Eprint
  {http://arxiv.org/abs/2010.07965} {arXiv:2010.07965 [quant-ph]} \BibitemShut
  {NoStop}%
\bibitem [{\citenamefont {Arute}\ \emph
  {et~al.}(2019{\natexlab{b}})\citenamefont {Arute}, \citenamefont {Arya},
  \citenamefont {Babbush}, \citenamefont {Bacon}, \citenamefont {Bardin},
  \citenamefont {Barends}, \citenamefont {Biswas}, \citenamefont {Boixo},
  \citenamefont {Brandao}, \citenamefont {Buell},\ and\ \citenamefont
  {et~al.}}]{2019}%
  \BibitemOpen
  \bibfield  {author} {\bibinfo {author} {\bibfnamefont {Frank}\ \bibnamefont
  {Arute}}, \bibinfo {author} {\bibfnamefont {Kunal}\ \bibnamefont {Arya}},
  \bibinfo {author} {\bibfnamefont {Ryan}\ \bibnamefont {Babbush}}, \bibinfo
  {author} {\bibfnamefont {Dave}\ \bibnamefont {Bacon}}, \bibinfo {author}
  {\bibfnamefont {Joseph~C.}\ \bibnamefont {Bardin}}, \bibinfo {author}
  {\bibfnamefont {Rami}\ \bibnamefont {Barends}}, \bibinfo {author}
  {\bibfnamefont {Rupak}\ \bibnamefont {Biswas}}, \bibinfo {author}
  {\bibfnamefont {Sergio}\ \bibnamefont {Boixo}}, \bibinfo {author}
  {\bibfnamefont {Fernando G. S.~L.}\ \bibnamefont {Brandao}}, \bibinfo
  {author} {\bibfnamefont {David~A.}\ \bibnamefont {Buell}}, \ and\ \bibinfo
  {author} {\bibnamefont {et~al.}},\ }\bibfield  {title} {\enquote {\bibinfo
  {title} {Quantum supremacy using a programmable superconducting processor},}\
  }\href {\doibase 10.1038/s41586-019-1666-5} {\bibfield  {journal} {\bibinfo
  {journal} {Nature}\ }\textbf {\bibinfo {volume} {574}},\ \bibinfo {pages}
  {505–510} (\bibinfo {year} {2019}{\natexlab{b}})}\BibitemShut {NoStop}%
\bibitem [{\citenamefont {Harrigan}\ \emph {et~al.}(2021)\citenamefont
  {Harrigan}, \citenamefont {Sung}, \citenamefont {Neeley}, \citenamefont
  {Satzinger}, \citenamefont {Arute}, \citenamefont {Arya}, \citenamefont
  {Atalaya}, \citenamefont {Bardin}, \citenamefont {Barends}, \citenamefont
  {Boixo}, \citenamefont {Broughton}, \citenamefont {Buckley}, \citenamefont
  {Buell}, \citenamefont {Burkett}, \citenamefont {Bushnell}, \citenamefont
  {Chen}, \citenamefont {Chen}, \citenamefont {Chiaro}, \citenamefont
  {Collins}, \citenamefont {Courtney}, \citenamefont {Demura}, \citenamefont
  {Dunsworth}, \citenamefont {Eppens}, \citenamefont {Fowler}, \citenamefont
  {Foxen}, \citenamefont {Gidney}, \citenamefont {Giustina}, \citenamefont
  {Graff}, \citenamefont {Habegger}, \citenamefont {Ho}, \citenamefont {Hong},
  \citenamefont {Huang}, \citenamefont {Ioffe}, \citenamefont {Isakov},
  \citenamefont {Jeffrey}, \citenamefont {Jiang}, \citenamefont {Jones},
  \citenamefont {Kafri}, \citenamefont {Kechedzhi}, \citenamefont {Kelly},
  \citenamefont {Kim}, \citenamefont {Klimov}, \citenamefont {Korotkov},
  \citenamefont {Kostritsa}, \citenamefont {Landhuis}, \citenamefont {Laptev},
  \citenamefont {Lindmark}, \citenamefont {Leib}, \citenamefont {Martin},
  \citenamefont {Martinis}, \citenamefont {McClean}, \citenamefont {McEwen},
  \citenamefont {Megrant}, \citenamefont {Mi}, \citenamefont {Mohseni},
  \citenamefont {Mruczkiewicz}, \citenamefont {Mutus}, \citenamefont {Naaman},
  \citenamefont {Neill}, \citenamefont {Neukart}, \citenamefont {Niu},
  \citenamefont {O'Brien}, \citenamefont {O'Gorman}, \citenamefont {Ostby},
  \citenamefont {Petukhov}, \citenamefont {Putterman}, \citenamefont
  {Quintana}, \citenamefont {Roushan}, \citenamefont {Rubin}, \citenamefont
  {Sank}, \citenamefont {Skolik}, \citenamefont {Smelyanskiy}, \citenamefont
  {Strain}, \citenamefont {Streif}, \citenamefont {Szalay}, \citenamefont
  {Vainsencher}, \citenamefont {White}, \citenamefont {Yao}, \citenamefont
  {Yeh}, \citenamefont {Zalcman}, \citenamefont {Zhou}, \citenamefont {Neven},
  \citenamefont {Bacon}, \citenamefont {Lucero}, \citenamefont {Farhi},\ and\
  \citenamefont {Babbush}}]{harrigan2021}%
  \BibitemOpen
  \bibfield  {author} {\bibinfo {author} {\bibfnamefont {Matthew~P.}\
  \bibnamefont {Harrigan}}, \bibinfo {author} {\bibfnamefont {Kevin~J.}\
  \bibnamefont {Sung}}, \bibinfo {author} {\bibfnamefont {Matthew}\
  \bibnamefont {Neeley}}, \bibinfo {author} {\bibfnamefont {Kevin~J.}\
  \bibnamefont {Satzinger}}, \bibinfo {author} {\bibfnamefont {Frank}\
  \bibnamefont {Arute}}, \bibinfo {author} {\bibfnamefont {Kunal}\ \bibnamefont
  {Arya}}, \bibinfo {author} {\bibfnamefont {Juan}\ \bibnamefont {Atalaya}},
  \bibinfo {author} {\bibfnamefont {Joseph~C.}\ \bibnamefont {Bardin}},
  \bibinfo {author} {\bibfnamefont {Rami}\ \bibnamefont {Barends}}, \bibinfo
  {author} {\bibfnamefont {Sergio}\ \bibnamefont {Boixo}}, \bibinfo {author}
  {\bibfnamefont {Michael}\ \bibnamefont {Broughton}}, \bibinfo {author}
  {\bibfnamefont {Bob~B.}\ \bibnamefont {Buckley}}, \bibinfo {author}
  {\bibfnamefont {David~A.}\ \bibnamefont {Buell}}, \bibinfo {author}
  {\bibfnamefont {Brian}\ \bibnamefont {Burkett}}, \bibinfo {author}
  {\bibfnamefont {Nicholas}\ \bibnamefont {Bushnell}}, \bibinfo {author}
  {\bibfnamefont {Yu}~\bibnamefont {Chen}}, \bibinfo {author} {\bibfnamefont
  {Zijun}\ \bibnamefont {Chen}}, \bibinfo {author} {\bibfnamefont {Ben}\
  \bibnamefont {Chiaro}}, \bibinfo {author} {\bibfnamefont {Roberto}\
  \bibnamefont {Collins}}, \bibinfo {author} {\bibfnamefont {William}\
  \bibnamefont {Courtney}}, \bibinfo {author} {\bibfnamefont {Sean}\
  \bibnamefont {Demura}}, \bibinfo {author} {\bibfnamefont {Andrew}\
  \bibnamefont {Dunsworth}}, \bibinfo {author} {\bibfnamefont {Daniel}\
  \bibnamefont {Eppens}}, \bibinfo {author} {\bibfnamefont {Austin}\
  \bibnamefont {Fowler}}, \bibinfo {author} {\bibfnamefont {Brooks}\
  \bibnamefont {Foxen}}, \bibinfo {author} {\bibfnamefont {Craig}\ \bibnamefont
  {Gidney}}, \bibinfo {author} {\bibfnamefont {Marissa}\ \bibnamefont
  {Giustina}}, \bibinfo {author} {\bibfnamefont {Rob}\ \bibnamefont {Graff}},
  \bibinfo {author} {\bibfnamefont {Steve}\ \bibnamefont {Habegger}}, \bibinfo
  {author} {\bibfnamefont {Alan}\ \bibnamefont {Ho}}, \bibinfo {author}
  {\bibfnamefont {Sabrina}\ \bibnamefont {Hong}}, \bibinfo {author}
  {\bibfnamefont {Trent}\ \bibnamefont {Huang}}, \bibinfo {author}
  {\bibfnamefont {L.~B.}\ \bibnamefont {Ioffe}}, \bibinfo {author}
  {\bibfnamefont {Sergei~V.}\ \bibnamefont {Isakov}}, \bibinfo {author}
  {\bibfnamefont {Evan}\ \bibnamefont {Jeffrey}}, \bibinfo {author}
  {\bibfnamefont {Zhang}\ \bibnamefont {Jiang}}, \bibinfo {author}
  {\bibfnamefont {Cody}\ \bibnamefont {Jones}}, \bibinfo {author}
  {\bibfnamefont {Dvir}\ \bibnamefont {Kafri}}, \bibinfo {author}
  {\bibfnamefont {Kostyantyn}\ \bibnamefont {Kechedzhi}}, \bibinfo {author}
  {\bibfnamefont {Julian}\ \bibnamefont {Kelly}}, \bibinfo {author}
  {\bibfnamefont {Seon}\ \bibnamefont {Kim}}, \bibinfo {author} {\bibfnamefont
  {Paul~V.}\ \bibnamefont {Klimov}}, \bibinfo {author} {\bibfnamefont
  {Alexander~N.}\ \bibnamefont {Korotkov}}, \bibinfo {author} {\bibfnamefont
  {Fedor}\ \bibnamefont {Kostritsa}}, \bibinfo {author} {\bibfnamefont {David}\
  \bibnamefont {Landhuis}}, \bibinfo {author} {\bibfnamefont {Pavel}\
  \bibnamefont {Laptev}}, \bibinfo {author} {\bibfnamefont {Mike}\ \bibnamefont
  {Lindmark}}, \bibinfo {author} {\bibfnamefont {Martin}\ \bibnamefont {Leib}},
  \bibinfo {author} {\bibfnamefont {Orion}\ \bibnamefont {Martin}}, \bibinfo
  {author} {\bibfnamefont {John~M.}\ \bibnamefont {Martinis}}, \bibinfo
  {author} {\bibfnamefont {Jarrod~R.}\ \bibnamefont {McClean}}, \bibinfo
  {author} {\bibfnamefont {Matt}\ \bibnamefont {McEwen}}, \bibinfo {author}
  {\bibfnamefont {Anthony}\ \bibnamefont {Megrant}}, \bibinfo {author}
  {\bibfnamefont {Xiao}\ \bibnamefont {Mi}}, \bibinfo {author} {\bibfnamefont
  {Masoud}\ \bibnamefont {Mohseni}}, \bibinfo {author} {\bibfnamefont
  {Wojciech}\ \bibnamefont {Mruczkiewicz}}, \bibinfo {author} {\bibfnamefont
  {Josh}\ \bibnamefont {Mutus}}, \bibinfo {author} {\bibfnamefont {Ofer}\
  \bibnamefont {Naaman}}, \bibinfo {author} {\bibfnamefont {Charles}\
  \bibnamefont {Neill}}, \bibinfo {author} {\bibfnamefont {Florian}\
  \bibnamefont {Neukart}}, \bibinfo {author} {\bibfnamefont {Murphy~Yuezhen}\
  \bibnamefont {Niu}}, \bibinfo {author} {\bibfnamefont {Thomas~E.}\
  \bibnamefont {O'Brien}}, \bibinfo {author} {\bibfnamefont {Bryan}\
  \bibnamefont {O'Gorman}}, \bibinfo {author} {\bibfnamefont {Eric}\
  \bibnamefont {Ostby}}, \bibinfo {author} {\bibfnamefont {Andre}\ \bibnamefont
  {Petukhov}}, \bibinfo {author} {\bibfnamefont {Harald}\ \bibnamefont
  {Putterman}}, \bibinfo {author} {\bibfnamefont {Chris}\ \bibnamefont
  {Quintana}}, \bibinfo {author} {\bibfnamefont {Pedram}\ \bibnamefont
  {Roushan}}, \bibinfo {author} {\bibfnamefont {Nicholas~C.}\ \bibnamefont
  {Rubin}}, \bibinfo {author} {\bibfnamefont {Daniel}\ \bibnamefont {Sank}},
  \bibinfo {author} {\bibfnamefont {Andrea}\ \bibnamefont {Skolik}}, \bibinfo
  {author} {\bibfnamefont {Vadim}\ \bibnamefont {Smelyanskiy}}, \bibinfo
  {author} {\bibfnamefont {Doug}\ \bibnamefont {Strain}}, \bibinfo {author}
  {\bibfnamefont {Michael}\ \bibnamefont {Streif}}, \bibinfo {author}
  {\bibfnamefont {Marco}\ \bibnamefont {Szalay}}, \bibinfo {author}
  {\bibfnamefont {Amit}\ \bibnamefont {Vainsencher}}, \bibinfo {author}
  {\bibfnamefont {Theodore}\ \bibnamefont {White}}, \bibinfo {author}
  {\bibfnamefont {Z.~Jamie}\ \bibnamefont {Yao}}, \bibinfo {author}
  {\bibfnamefont {Ping}\ \bibnamefont {Yeh}}, \bibinfo {author} {\bibfnamefont
  {Adam}\ \bibnamefont {Zalcman}}, \bibinfo {author} {\bibfnamefont {Leo}\
  \bibnamefont {Zhou}}, \bibinfo {author} {\bibfnamefont {Hartmut}\
  \bibnamefont {Neven}}, \bibinfo {author} {\bibfnamefont {Dave}\ \bibnamefont
  {Bacon}}, \bibinfo {author} {\bibfnamefont {Erik}\ \bibnamefont {Lucero}},
  \bibinfo {author} {\bibfnamefont {Edward}\ \bibnamefont {Farhi}}, \ and\
  \bibinfo {author} {\bibfnamefont {Ryan}\ \bibnamefont {Babbush}},\ }\bibfield
   {title} {\enquote {\bibinfo {title} {Quantum approximate optimization of
  non-planar graph problems on a planar superconducting processor},}\ }\href
  {\doibase 10.1038/s41567-020-01105-y} {\ \textbf {\bibinfo {volume} {17}},\
  \bibinfo {pages} {332--336} (\bibinfo {year} {2021})}\BibitemShut {NoStop}%
\bibitem [{\citenamefont {Kivlichan}\ \emph {et~al.}(2018)\citenamefont
  {Kivlichan}, \citenamefont {McClean}, \citenamefont {Wiebe}, \citenamefont
  {Gidney}, \citenamefont {Aspuru-Guzik}, \citenamefont {Chan},\ and\
  \citenamefont {Babbush}}]{PhysRevLett.120.110501}%
  \BibitemOpen
  \bibfield  {author} {\bibinfo {author} {\bibfnamefont {Ian~D.}\ \bibnamefont
  {Kivlichan}}, \bibinfo {author} {\bibfnamefont {Jarrod}\ \bibnamefont
  {McClean}}, \bibinfo {author} {\bibfnamefont {Nathan}\ \bibnamefont {Wiebe}},
  \bibinfo {author} {\bibfnamefont {Craig}\ \bibnamefont {Gidney}}, \bibinfo
  {author} {\bibfnamefont {Al\'an}\ \bibnamefont {Aspuru-Guzik}}, \bibinfo
  {author} {\bibfnamefont {Garnet Kin-Lic}\ \bibnamefont {Chan}}, \ and\
  \bibinfo {author} {\bibfnamefont {Ryan}\ \bibnamefont {Babbush}},\ }\bibfield
   {title} {\enquote {\bibinfo {title} {Quantum simulation of electronic
  structure with linear depth and connectivity},}\ }\href {\doibase
  10.1103/PhysRevLett.120.110501} {\bibfield  {journal} {\bibinfo  {journal}
  {Phys. Rev. Lett.}\ }\textbf {\bibinfo {volume} {120}},\ \bibinfo {pages}
  {110501} (\bibinfo {year} {2018})}\BibitemShut {NoStop}%
\bibitem [{\citenamefont {Arute}\ \emph
  {et~al.}(2020{\natexlab{b}})\citenamefont {Arute}, \citenamefont {Arya},
  \citenamefont {Babbush}, \citenamefont {Bacon}, \citenamefont {Bardin},
  \citenamefont {Barends}, \citenamefont {Boixo}, \citenamefont {Broughton},
  \citenamefont {Buckley}, \citenamefont {Buell}, \citenamefont {Burkett},
  \citenamefont {Bushnell}, \citenamefont {Chen}, \citenamefont {Chen},
  \citenamefont {Chiaro}, \citenamefont {Collins}, \citenamefont {Courtney},
  \citenamefont {Demura}, \citenamefont {Dunsworth}, \citenamefont {Farhi},
  \citenamefont {Fowler}, \citenamefont {Foxen}, \citenamefont {Gidney},
  \citenamefont {Giustina}, \citenamefont {Graff}, \citenamefont {Habegger},
  \citenamefont {Harrigan}, \citenamefont {Ho}, \citenamefont {Hong},
  \citenamefont {Huang}, \citenamefont {Huggins}, \citenamefont {Ioffe},
  \citenamefont {Isakov}, \citenamefont {Jeffrey}, \citenamefont {Jiang},
  \citenamefont {Jones}, \citenamefont {Kafri}, \citenamefont {Kechedzhi},
  \citenamefont {Kelly}, \citenamefont {Kim}, \citenamefont {Klimov},
  \citenamefont {Korotkov}, \citenamefont {Kostritsa}, \citenamefont
  {Landhuis}, \citenamefont {Laptev}, \citenamefont {Lindmark}, \citenamefont
  {Lucero}, \citenamefont {Martin}, \citenamefont {Martinis}, \citenamefont
  {McClean}, \citenamefont {McEwen}, \citenamefont {Megrant}, \citenamefont
  {Mi}, \citenamefont {Mohseni}, \citenamefont {Mruczkiewicz}, \citenamefont
  {Mutus}, \citenamefont {Naaman}, \citenamefont {Neeley}, \citenamefont
  {Neill}, \citenamefont {Neven}, \citenamefont {Niu}, \citenamefont
  {O’Brien}, \citenamefont {Ostby}, \citenamefont {Petukhov}, \citenamefont
  {Putterman}, \citenamefont {Quintana}, \citenamefont {Roushan}, \citenamefont
  {Rubin}, \citenamefont {Sank}, \citenamefont {Satzinger}, \citenamefont
  {Smelyanskiy}, \citenamefont {Strain}, \citenamefont {Sung}, \citenamefont
  {Szalay}, \citenamefont {Takeshita}, \citenamefont {Vainsencher},
  \citenamefont {White}, \citenamefont {Wiebe}, \citenamefont {Yao},
  \citenamefont {Yeh},\ and\ \citenamefont {Zalcman}}]{Rubinscience.abb9811}%
  \BibitemOpen
  \bibfield  {author} {\bibinfo {author} {\bibfnamefont {Frank}\ \bibnamefont
  {Arute}}, \bibinfo {author} {\bibfnamefont {Kunal}\ \bibnamefont {Arya}},
  \bibinfo {author} {\bibfnamefont {Ryan}\ \bibnamefont {Babbush}}, \bibinfo
  {author} {\bibfnamefont {Dave}\ \bibnamefont {Bacon}}, \bibinfo {author}
  {\bibfnamefont {Joseph~C.}\ \bibnamefont {Bardin}}, \bibinfo {author}
  {\bibfnamefont {Rami}\ \bibnamefont {Barends}}, \bibinfo {author}
  {\bibfnamefont {Sergio}\ \bibnamefont {Boixo}}, \bibinfo {author}
  {\bibfnamefont {Michael}\ \bibnamefont {Broughton}}, \bibinfo {author}
  {\bibfnamefont {Bob~B.}\ \bibnamefont {Buckley}}, \bibinfo {author}
  {\bibfnamefont {David~A.}\ \bibnamefont {Buell}}, \bibinfo {author}
  {\bibfnamefont {Brian}\ \bibnamefont {Burkett}}, \bibinfo {author}
  {\bibfnamefont {Nicholas}\ \bibnamefont {Bushnell}}, \bibinfo {author}
  {\bibfnamefont {Yu}~\bibnamefont {Chen}}, \bibinfo {author} {\bibfnamefont
  {Zijun}\ \bibnamefont {Chen}}, \bibinfo {author} {\bibfnamefont {Benjamin}\
  \bibnamefont {Chiaro}}, \bibinfo {author} {\bibfnamefont {Roberto}\
  \bibnamefont {Collins}}, \bibinfo {author} {\bibfnamefont {William}\
  \bibnamefont {Courtney}}, \bibinfo {author} {\bibfnamefont {Sean}\
  \bibnamefont {Demura}}, \bibinfo {author} {\bibfnamefont {Andrew}\
  \bibnamefont {Dunsworth}}, \bibinfo {author} {\bibfnamefont {Edward}\
  \bibnamefont {Farhi}}, \bibinfo {author} {\bibfnamefont {Austin}\
  \bibnamefont {Fowler}}, \bibinfo {author} {\bibfnamefont {Brooks}\
  \bibnamefont {Foxen}}, \bibinfo {author} {\bibfnamefont {Craig}\ \bibnamefont
  {Gidney}}, \bibinfo {author} {\bibfnamefont {Marissa}\ \bibnamefont
  {Giustina}}, \bibinfo {author} {\bibfnamefont {Rob}\ \bibnamefont {Graff}},
  \bibinfo {author} {\bibfnamefont {Steve}\ \bibnamefont {Habegger}}, \bibinfo
  {author} {\bibfnamefont {Matthew~P.}\ \bibnamefont {Harrigan}}, \bibinfo
  {author} {\bibfnamefont {Alan}\ \bibnamefont {Ho}}, \bibinfo {author}
  {\bibfnamefont {Sabrina}\ \bibnamefont {Hong}}, \bibinfo {author}
  {\bibfnamefont {Trent}\ \bibnamefont {Huang}}, \bibinfo {author}
  {\bibfnamefont {William~J.}\ \bibnamefont {Huggins}}, \bibinfo {author}
  {\bibfnamefont {Lev}\ \bibnamefont {Ioffe}}, \bibinfo {author} {\bibfnamefont
  {Sergei~V.}\ \bibnamefont {Isakov}}, \bibinfo {author} {\bibfnamefont {Evan}\
  \bibnamefont {Jeffrey}}, \bibinfo {author} {\bibfnamefont {Zhang}\
  \bibnamefont {Jiang}}, \bibinfo {author} {\bibfnamefont {Cody}\ \bibnamefont
  {Jones}}, \bibinfo {author} {\bibfnamefont {Dvir}\ \bibnamefont {Kafri}},
  \bibinfo {author} {\bibfnamefont {Kostyantyn}\ \bibnamefont {Kechedzhi}},
  \bibinfo {author} {\bibfnamefont {Julian}\ \bibnamefont {Kelly}}, \bibinfo
  {author} {\bibfnamefont {Seon}\ \bibnamefont {Kim}}, \bibinfo {author}
  {\bibfnamefont {Paul~V.}\ \bibnamefont {Klimov}}, \bibinfo {author}
  {\bibfnamefont {Alexander}\ \bibnamefont {Korotkov}}, \bibinfo {author}
  {\bibfnamefont {Fedor}\ \bibnamefont {Kostritsa}}, \bibinfo {author}
  {\bibfnamefont {David}\ \bibnamefont {Landhuis}}, \bibinfo {author}
  {\bibfnamefont {Pavel}\ \bibnamefont {Laptev}}, \bibinfo {author}
  {\bibfnamefont {Mike}\ \bibnamefont {Lindmark}}, \bibinfo {author}
  {\bibfnamefont {Erik}\ \bibnamefont {Lucero}}, \bibinfo {author}
  {\bibfnamefont {Orion}\ \bibnamefont {Martin}}, \bibinfo {author}
  {\bibfnamefont {John~M.}\ \bibnamefont {Martinis}}, \bibinfo {author}
  {\bibfnamefont {Jarrod~R.}\ \bibnamefont {McClean}}, \bibinfo {author}
  {\bibfnamefont {Matt}\ \bibnamefont {McEwen}}, \bibinfo {author}
  {\bibfnamefont {Anthony}\ \bibnamefont {Megrant}}, \bibinfo {author}
  {\bibfnamefont {Xiao}\ \bibnamefont {Mi}}, \bibinfo {author} {\bibfnamefont
  {Masoud}\ \bibnamefont {Mohseni}}, \bibinfo {author} {\bibfnamefont
  {Wojciech}\ \bibnamefont {Mruczkiewicz}}, \bibinfo {author} {\bibfnamefont
  {Josh}\ \bibnamefont {Mutus}}, \bibinfo {author} {\bibfnamefont {Ofer}\
  \bibnamefont {Naaman}}, \bibinfo {author} {\bibfnamefont {Matthew}\
  \bibnamefont {Neeley}}, \bibinfo {author} {\bibfnamefont {Charles}\
  \bibnamefont {Neill}}, \bibinfo {author} {\bibfnamefont {Hartmut}\
  \bibnamefont {Neven}}, \bibinfo {author} {\bibfnamefont {Murphy~Yuezhen}\
  \bibnamefont {Niu}}, \bibinfo {author} {\bibfnamefont {Thomas~E.}\
  \bibnamefont {O’Brien}}, \bibinfo {author} {\bibfnamefont {Eric}\
  \bibnamefont {Ostby}}, \bibinfo {author} {\bibfnamefont {Andre}\ \bibnamefont
  {Petukhov}}, \bibinfo {author} {\bibfnamefont {Harald}\ \bibnamefont
  {Putterman}}, \bibinfo {author} {\bibfnamefont {Chris}\ \bibnamefont
  {Quintana}}, \bibinfo {author} {\bibfnamefont {Pedram}\ \bibnamefont
  {Roushan}}, \bibinfo {author} {\bibfnamefont {Nicholas~C.}\ \bibnamefont
  {Rubin}}, \bibinfo {author} {\bibfnamefont {Daniel}\ \bibnamefont {Sank}},
  \bibinfo {author} {\bibfnamefont {Kevin~J.}\ \bibnamefont {Satzinger}},
  \bibinfo {author} {\bibfnamefont {Vadim}\ \bibnamefont {Smelyanskiy}},
  \bibinfo {author} {\bibfnamefont {Doug}\ \bibnamefont {Strain}}, \bibinfo
  {author} {\bibfnamefont {Kevin~J.}\ \bibnamefont {Sung}}, \bibinfo {author}
  {\bibfnamefont {Marco}\ \bibnamefont {Szalay}}, \bibinfo {author}
  {\bibfnamefont {Tyler~Y.}\ \bibnamefont {Takeshita}}, \bibinfo {author}
  {\bibfnamefont {Amit}\ \bibnamefont {Vainsencher}}, \bibinfo {author}
  {\bibfnamefont {Theodore}\ \bibnamefont {White}}, \bibinfo {author}
  {\bibfnamefont {Nathan}\ \bibnamefont {Wiebe}}, \bibinfo {author}
  {\bibfnamefont {Z.~Jamie}\ \bibnamefont {Yao}}, \bibinfo {author}
  {\bibfnamefont {Ping}\ \bibnamefont {Yeh}}, \ and\ \bibinfo {author}
  {\bibfnamefont {Adam}\ \bibnamefont {Zalcman}},\ }\bibfield  {title}
  {\enquote {\bibinfo {title} {Hartree-fock on a superconducting qubit quantum
  computer},}\ }\href {\doibase 10.1126/science.abb9811} {\bibfield  {journal}
  {\bibinfo  {journal} {Science}\ }\textbf {\bibinfo {volume} {369}},\ \bibinfo
  {pages} {1084--1089} (\bibinfo {year} {2020}{\natexlab{b}})}\BibitemShut
  {NoStop}%
\bibitem [{\citenamefont {Jiang}\ \emph {et~al.}(2018)\citenamefont {Jiang},
  \citenamefont {Sung}, \citenamefont {Kechedzhi}, \citenamefont
  {Smelyanskiy},\ and\ \citenamefont {Boixo}}]{PhysRevApplied.9.044036}%
  \BibitemOpen
  \bibfield  {author} {\bibinfo {author} {\bibfnamefont {Zhang}\ \bibnamefont
  {Jiang}}, \bibinfo {author} {\bibfnamefont {Kevin~J.}\ \bibnamefont {Sung}},
  \bibinfo {author} {\bibfnamefont {Kostyantyn}\ \bibnamefont {Kechedzhi}},
  \bibinfo {author} {\bibfnamefont {Vadim~N.}\ \bibnamefont {Smelyanskiy}}, \
  and\ \bibinfo {author} {\bibfnamefont {Sergio}\ \bibnamefont {Boixo}},\
  }\bibfield  {title} {\enquote {\bibinfo {title} {Quantum algorithms to
  simulate many-body physics of correlated fermions},}\ }\href {\doibase
  10.1103/PhysRevApplied.9.044036} {\bibfield  {journal} {\bibinfo  {journal}
  {Phys. Rev. Applied}\ }\textbf {\bibinfo {volume} {9}},\ \bibinfo {pages}
  {044036} (\bibinfo {year} {2018})}\BibitemShut {NoStop}%
\bibitem [{\citenamefont {Rubin}\ \emph {et~al.}(2021)\citenamefont {Rubin},
  \citenamefont {Lee},\ and\ \citenamefont {Babbush}}]{rubin2021compressing}%
  \BibitemOpen
  \bibfield  {author} {\bibinfo {author} {\bibfnamefont {Nicholas~C}\
  \bibnamefont {Rubin}}, \bibinfo {author} {\bibfnamefont {Joonho}\
  \bibnamefont {Lee}}, \ and\ \bibinfo {author} {\bibfnamefont {Ryan}\
  \bibnamefont {Babbush}},\ }\bibfield  {title} {\enquote {\bibinfo {title}
  {Compressing many-body fermion operators under unitary constraints},}\ }\href
  {https://arxiv.org/abs/2109.05010} {\  (\bibinfo {year} {2021})},\ \Eprint
  {http://arxiv.org/abs/2109.05010} {arXiv:2109.05010 [quant-ph]} \BibitemShut
  {NoStop}%
\bibitem [{\citenamefont {Jozsa}\ and\ \citenamefont
  {Miyake}(2008)}]{jozsa2008matchgates}%
  \BibitemOpen
  \bibfield  {author} {\bibinfo {author} {\bibfnamefont {Richard}\ \bibnamefont
  {Jozsa}}\ and\ \bibinfo {author} {\bibfnamefont {Akimasa}\ \bibnamefont
  {Miyake}},\ }\bibfield  {title} {\enquote {\bibinfo {title} {Matchgates and
  classical simulation of quantum circuits},}\ }\href@noop {} {\bibfield
  {journal} {\bibinfo  {journal} {Proceedings of the Royal Society A:
  Mathematical, Physical and Engineering Sciences}\ }\textbf {\bibinfo {volume}
  {464}},\ \bibinfo {pages} {3089--3106} (\bibinfo {year} {2008})}\BibitemShut
  {NoStop}%
\bibitem [{\citenamefont {Clements}\ \emph {et~al.}(2016)\citenamefont
  {Clements}, \citenamefont {Humphreys}, \citenamefont {Metcalf}, \citenamefont
  {Kolthammer},\ and\ \citenamefont {Walmsley}}]{clements2016optimal}%
  \BibitemOpen
  \bibfield  {author} {\bibinfo {author} {\bibfnamefont {William~R}\
  \bibnamefont {Clements}}, \bibinfo {author} {\bibfnamefont {Peter~C}\
  \bibnamefont {Humphreys}}, \bibinfo {author} {\bibfnamefont {Benjamin~J}\
  \bibnamefont {Metcalf}}, \bibinfo {author} {\bibfnamefont {W~Steven}\
  \bibnamefont {Kolthammer}}, \ and\ \bibinfo {author} {\bibfnamefont {Ian~A}\
  \bibnamefont {Walmsley}},\ }\bibfield  {title} {\enquote {\bibinfo {title}
  {Optimal design for universal multiport interferometers},}\ }\href
  {https://www.osapublishing.org/optica/fulltext.cfm?uri=optica-3-12-1460&id=355743}
  {\bibfield  {journal} {\bibinfo  {journal} {Optica}\ }\textbf {\bibinfo
  {volume} {3}},\ \bibinfo {pages} {1460--1465} (\bibinfo {year}
  {2016})}\BibitemShut {NoStop}%
\bibitem [{\citenamefont {Reck}\ \emph {et~al.}(1994)\citenamefont {Reck},
  \citenamefont {Zeilinger}, \citenamefont {Bernstein},\ and\ \citenamefont
  {Bertani}}]{PhysRevLett.73.58}%
  \BibitemOpen
  \bibfield  {author} {\bibinfo {author} {\bibfnamefont {Michael}\ \bibnamefont
  {Reck}}, \bibinfo {author} {\bibfnamefont {Anton}\ \bibnamefont {Zeilinger}},
  \bibinfo {author} {\bibfnamefont {Herbert~J.}\ \bibnamefont {Bernstein}}, \
  and\ \bibinfo {author} {\bibfnamefont {Philip}\ \bibnamefont {Bertani}},\
  }\bibfield  {title} {\enquote {\bibinfo {title} {Experimental realization of
  any discrete unitary operator},}\ }\href {\doibase 10.1103/PhysRevLett.73.58}
  {\bibfield  {journal} {\bibinfo  {journal} {Phys. Rev. Lett.}\ }\textbf
  {\bibinfo {volume} {73}},\ \bibinfo {pages} {58--61} (\bibinfo {year}
  {1994})}\BibitemShut {NoStop}%
\end{thebibliography}%

\end{document}